\documentclass[fleqn]{article}
\usepackage[margin=1.2in]{geometry} %
\usepackage{mathtools}
\usepackage{amssymb}
\usepackage{array}
\usepackage{booktabs}
\usepackage{mathrsfs}
\usepackage{dsfont}
\usepackage{cite}%
\usepackage{graphicx}
\usepackage{xcolor}
\usepackage{nicefrac}
\usepackage[hidelinks]{hyperref}
\usepackage[noabbrev]{cleveref}
\usepackage{float}
\usepackage[nolist]{acronym}
\usepackage{xspace}
\usepackage[small]{caption}
\usepackage{mleftright}
\usepackage[symbol]{footmisc}
\mleftright
\makeatletter
\g@addto@macro\bfseries{\boldmath}
\makeatother
\newcommand{\rep}[1]{\ensuremath{\boldsymbol{#1}}}
\newcommand{\crep}[1]{\ensuremath{\overline{\boldsymbol{#1}}}}
\DeclareMathOperator{\re}{Re}
\DeclareMathOperator{\im}{Im}

\DeclareMathOperator{\diag}{diag}

\DeclareMathOperator{\lcm}{lcm}
\newcommand{\D}{\mathop{}\!\mathrm{d}}
\newcommand{\I}{\hskip0.1ex\mathrm{i}\hskip0.1ex}
\newcommand{\Id}{\ensuremath{\mathds{1}}}
\newcommand{\SU}[1]{\ensuremath{\mathrm{SU}(#1)}}
\newcommand{\SL}[1]{\ensuremath{\mathrm{SL}(#1)}}
\newcommand{\U}[1]{\ensuremath{\mathrm{U}(#1)}}
\newcommand{\Z}[1]{\ensuremath{\mathds{Z}_{#1}}} 
\newcommand{\CP}{\ensuremath{\mathcal{CP}}\xspace}
\newcommand{\orb}[2]{\hspace*{-0.1ex}\begin{bsmallmatrix}\!#1\!\\[0.3ex]\!#2\!\end{bsmallmatrix}\hspace*{-0.1ex}}
\newcommand{\Orb}[2]{\hspace*{-0.1ex}\begin{bmatrix}\hspace*{-0.1ex}#1\hspace*{-0.1ex}\\\hspace*{-0.1ex}#2\hspace*{-0.1ex}\end{bmatrix}\hspace*{-0.1ex}}
\newcommand{\wavefunction}[1][j,M]{\ensuremath{\psi^{#1}}}
\def\mytitle{Metaplectic Flavor Symmetries from Magnetized Tori}
\title{\mytitle}
\numberwithin{equation}{section}
\numberwithin{figure}{section}
\numberwithin{table}{section}
\begin{document}
\begin{titlepage}
\vspace*{1.0cm}

\begin{flushright}
UCI--TR--2021--08%
\\
\end{flushright}

\vspace*{2cm}

\begin{center}
{\LARGE\sffamily\bfseries\mytitle}

\vspace{1cm}

\renewcommand*{\thefootnote}{\ensuremath{\ifcase\value{footnote}%
\or\alpha\or\beta\or\gamma\or\delta\or\varepsilon\or\varphi\fi}}

\textbf{%
Yahya Almumin$^{a,}$\footnote{yalmumin@uci.edu}, Mu--Chun Chen$^{a,}$\footnote{muchunc@uci.edu}, 
V\'ictor Knapp--P\'erez$^{b,c,}$\footnote{victorknapp@ciencias.unam.mx},\\
Sa\'ul Ramos--S\'anchez$^{b,}$\footnote{ramos@fisica.unam.mx}, Michael Ratz$^{a,}$\footnote{mratz@uci.edu} 
and Shreya Shukla$^{a,}$\footnote{sshukla4@uci.edu}
}
\\[8mm]
\textit{$^a$\small
~Department of Physics and Astronomy, University of California, Irvine, CA 92697-4575 USA
}
\\[5mm]
\textit{$^b$\small Instituto de F\'isica, Universidad Nacional Aut\'onoma de M\'exico, POB 20-364, Cd.Mx. 01000, M\'exico}
\\[5mm]
\textit{$^c$\small Address from September 2021: Department of Physics and Astronomy, University of California, Irvine, CA 92697-4575 USA}
\end{center}

\vspace*{1cm}

\begin{abstract}
We revisit the flavor symmetries arising from compactifications on tori with
magnetic background fluxes. Using Euler's Theorem, we derive closed form
analytic expressions for the Yukawa couplings that are valid for arbitrary flux
parameters. We discuss the modular transformations for even and odd units of
magnetic flux, $M$, and show that they give rise to finite metaplectic groups the
order of which is determined by the least common multiple of the number of
zero--mode flavors involved. Unlike in models in which modular flavor symmetries
are postulated,  in this approach they derive from an underlying torus. This
allows us to retain control over parameters, such as those governing the kinetic
terms, that are free in the bottom--up approach, thus leading to an increased
predictivity. In addition, the geometric picture allows us to understand the
relative suppression of Yukawa couplings from their localization properties in
the compact space. We also comment on the role supersymmetry plays in these
constructions, and outline a path towards non--supersymmetric models with
modular flavor symmetries.
\end{abstract}
\vspace*{1cm}
\end{titlepage}
\renewcommand*{\thefootnote}{\arabic{footnote}}
\setcounter{footnote}{0}


\section{Introduction}
\label{sec:intro}

The \ac{SM} of particle physics is believed to be an effective theory. One
reason why this is so is that it has many parameters that have to be adjusted by
hand to fit data. The bulk of these parameters resides in the flavor sector,
i.e.\ concerns the fermion masses, mixing angles and \CP phases. An \ac{UV}
completion of the \ac{SM} will have to explain these parameters. Turning this
around, one may hope to get more insights on the \ac{UV} completion by
constructing a working theory of flavor. 

Recently, a new approach to address the flavor problem has been put forward
\cite{Feruglio:2017spp}: Yukawa couplings could be modular forms. 
There are two main ways in which this proposal has been utilized: 
\begin{enumerate}
 \item \emph{\ac{SB}}, i.e.\ impose the modular flavor symmetry to construct
  the Lagrange density \cite{
Kobayashi:2018vbk,           
Penedo:2018nmg,              
Criado:2018thu,              
deAnda:2018ecu,              
Okada:2018yrn,               
Ding:2019xna,                
Novichkov:2019sqv,           
Liu:2019khw,                 
Kobayashi:2019xvz,           
Asaka:2019vev,               
Gui-JunDing:2019wap,         
Kobayashi:2019uyt,           
Ding:2020yen,                
Liu:2020msy,                 
Ding:2020zxw,				 
Yao:2020zml
},
and
 \item \emph{\ac{TB}}, in which one derives the symmetries from an underlying
  torus or related setup 
  \cite{Kobayashi:2016ovu,
  	Kobayashi:2018rad,
	Kobayashi:2018bff,
	Kariyazono:2019ehj,
  	Baur:2019kwi,
	Nilles:2020tdp,
	Baur:2020jwc,
	Baur:2020yjl,
    Ohki:2020bpo,Kikuchi:2020frp,Kikuchi:2020nxn,Hoshiya:2020hki,Kikuchi:2021ogn}.
\end{enumerate}
Both strategies have strong points and challenges. In the \ac{SB} approach, very
good fits to data have been achieved. However, this is, in part, possible
because one can postulate the symmetry and other data like modular weights and
representations at will. Apart from the arbitrariness of the flavor group and
modular weights, the kinetic terms of the fields are not very constrained by the
modular transformations~\cite{Chen:2019ewa}. The \ac{TB} approach is much more
restrictive, in particular when embedded into string theory
\cite{Baur:2019kwi,Nilles:2020tdp,Baur:2020jwc,Baur:2020yjl}. However, while
these models have great promise and certainly fix the above--mentioned problems
of arbitrariness, it is probably fair to say that they do not yet provide us
with unequivocal predictions on flavor parameters that can be tested in the
foreseeable future.

The purpose of this paper is to explore the details of the relation between
these approaches. More specifically, we derive metaplectic symmetries from
magnetized tori. Earlier works on this subject include
\cite{Kobayashi:2016ovu,Kobayashi:2018rad,Kobayashi:2018bff,Kariyazono:2019ehj,Ohki:2020bpo,Kikuchi:2020frp,Kikuchi:2020nxn,Kikuchi:2021ogn}.
To accomplish this, we work out closed--form expressions for the Yukawa
couplings that are valid for arbitrary flux parameters, and thus generalize the
results of the pioneering work by Cremades, Ib\'a\~nez and
Marchesano~\cite{Cremades:2004wa}. We also present consistent modular
transformation laws for both even and odd numbers of generations. Models derived
from magnetized tori also allow us to understand to which extent supersymmetry
is crucial for modular flavor symmetries, which we will argue to be less
important than usually assumed. Additional motivation for looking at magnetized
tori, with and without supersymmetry, comes from the fact that even without
supersymmetry interacting scalar masses seem to be protected from quantum
corrections
\cite{Buchmuller:2016gib,Ghilencea:2017jmh,Buchmuller:2018eog,Hirose:2019ywp}.

This paper is organized as follows. In \cref{sec:zero-modes} we review the zero
modes on magetized tori. \Cref{sec:YukawaCouplings} concerns the computation of
the Yukawa couplings of these settings. We derive closed form expressions that
are valid for arbitrary flux parameters. In \cref{sec:ModularTransformations} we
show how modular transformations amount to flavor rotations. We will show that
the torus compactifications give rise to finite metaplectic groups, which have
been studied using the \ac{SB} approach in \cite{Liu:2020msy,Yao:2020zml}. In
\cref{sec:CommentsOnSUSY} we comment on the role that supersymmetry plays in the
scheme of modular flavor symmetries. \Cref{sec:Summary} contains our
conclusions. Various appendices contain some details of our derivations.

\section{Zero modes on tori with magnetic flux}
\label{sec:zero-modes}

Let us consider a gauge theory with two extra dimensions. The two extra
dimensions are compactified on a 2--torus $\mathds{T}^2$, which is endowed with a
magnetic flux. By the index theorem, the flux will give rise to chiral
zero--modes. Throughout our discussion we will ignore questions on the vacuum
energy, the stability and even anomaly cancellation. We think of this torus as a
little local playground that is embedded in a more complete setup. However, we
will address some of the questions in \cref{sec:CommentsOnSUSY}.

\begin{figure}[t!]
\centering\includegraphics{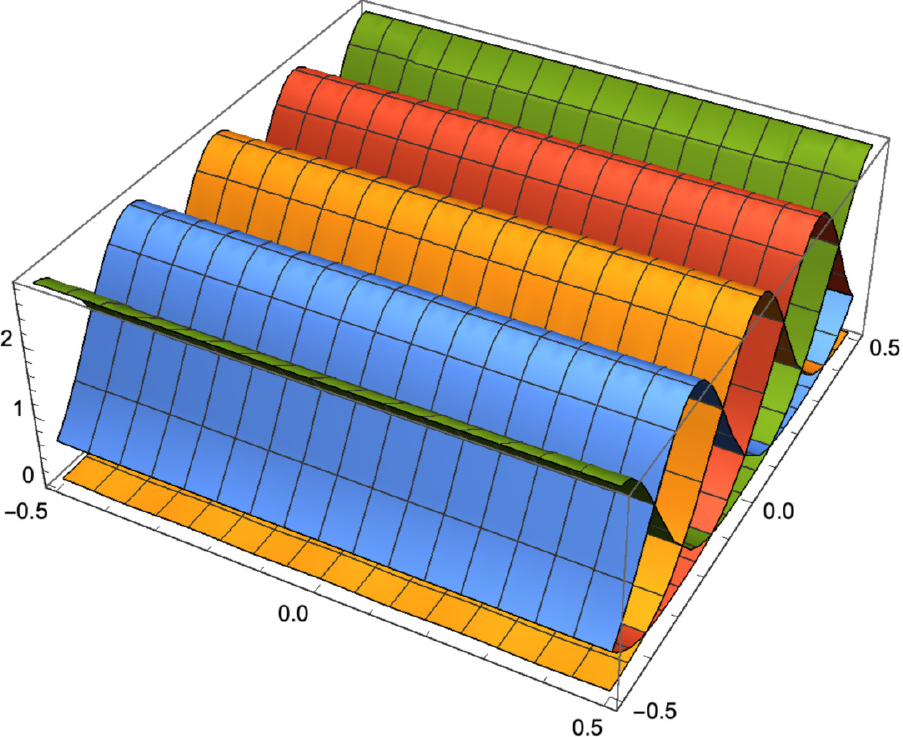}
\caption{Squares of the absolute values of the wave functions on a quadratic
torus for $M=4$.}
\label{fig:4wavefunctions}
\end{figure}

The main goal of this section is to review some of the properties of the
zero--modes. The wave functions of the zero modes of the Dirac operator on tori
with magnetic flux have been worked out in \cite{Cremades:2004wa}. They are
given by
\begin{align}
 \wavefunction(z,\tau,\zeta)&=\mathcal{N}\,
 \mathrm{e}^{\pi\,\I\,M\,(z+\zeta)\,\frac{\im (z+\zeta)}{\im\tau}}
 \,\vartheta\orb{\frac{j}{M}}{0}\bigl(M\,(z+\zeta),M\,\tau\bigr)\;.
 \label{eq:ZeroModesWavefunctions}
\end{align}
Here, $M\in\mathds{N}$ indicates the units of flux, $0\le j\le M-1$ is an integer, $z$
the coordinate in the extra dimensions, $\zeta$ a so--called Wilson line
parameter, and $\tau$ the torus parameter or half-period ratio. The wave functions from
\cref{eq:ZeroModesWavefunctions} correspond to left--handed particles in 4D
whereas there are no right--handed particles for positive $M$. On the other hand,
for negative values of the integer $M$ there are no solutions for left--handed
particles, but there are $|M|$ right--handed particles described by
$\wavefunction(\bar{z},\bar{\tau},\bar{\zeta })$, with $0\le j\le |M|-1$.
Furthermore, notice that despite what the notation may suggest, the
$\wavefunction$ are neither holomorphic functions of $z$, nor of $\tau$.
$\vartheta$ denotes the so--called Jacobi $\vartheta$--function, cf.\
\cref{sec:theta-functions}. The normalization is given by
\begin{align}\label{eq:Normalization}
 \mathcal{N}&=\left(\frac{2M\,\im\tau}{\mathcal{A}^{2}}\right)^{1/4}\;,
\end{align}
where $\mathcal{A}=(2\pi R)^{2}\text{Im}\tau$ is the area of the torus (cf.\
\cref{sec:Normalization}). In \cref{fig:4wavefunctions}, we show the profiles of
some zero--modes.

We find it instructive to derive the quantization condition on $M$.  Let us
follow the discussion by \cite{Cremades:2004wa}. Consider a $\U1$ gauge
group in the torus with a magnetic flux given by the gauge potential
\begin{equation}
    A(z+\zeta)=\frac{B}{2\im\tau}\im\bigl((\bar{z}+\bar{\zeta} )\D z \bigr)
	\;.
    \label{eq:Avectorpotential}
\end{equation}
Then, if the wave function $\wavefunction(z, \tau,\zeta)$ has charge $q$ under
this \U1, its transformation under torus translations are 
\begin{subequations}
\begin{align}
    \wavefunction(z+1,\tau,\zeta)&=\exp\bigg(\frac{\I\,qB}{2\im\tau}\im(z+\zeta)\bigg)
	\,\wavefunction(z,\tau,\zeta)
    \label{eq:boundaryqB1}\;,\\
    \wavefunction(z+\tau,\tau,\zeta)&=\exp\bigg(\frac{\I\,qB}{2\im\tau}\im(z+\zeta)\bar{\tau}\bigg)
	\,\wavefunction(z,\tau,\zeta)\;.
    \label{eq:boundaryqB2}
\end{align}
\end{subequations}
In order to have consistency through a contractible loop in the torus, we must
get the same wave function shifting $z\rightarrow z+\tau+1$ as in the case where
we shift by $z\rightarrow z+1+\tau$. Then,
\begin{align}
\MoveEqLeft\wavefunction(z+\tau+1,\tau,\zeta)=
 \exp\bigg(\frac{\I\,qB}{2\im\tau}\im\bigl((z+\zeta+1\bigr)\bar{\tau})\bigg)
 	\wavefunction(z+1,\tau,\zeta)\nonumber\\
 &=\exp\bigg(\frac{\I\,qB}{2\im\tau}\im\bigl((z+\zeta+1\bigr)\bar{\tau})\bigg)\exp\bigg(\frac{\I\,qB}{2\im\tau}\im(z+\zeta)\bigg)
 	\wavefunction(z,\tau,\zeta)\nonumber\\
 &=\exp\bigg(\I\,\frac{q B\im\bar{\tau}}{2\im\tau}\bigg)
  \exp\bigg(\frac{\I\,qB}{2\im\tau}\im\bigl(\bar{\tau}(z+\zeta)\bigr)\bigg)
  \exp\bigg(\frac{\I\,qB}{2\im\tau}\im(z+\zeta)\bigg)
  \wavefunction(z,\tau,\zeta)\mathrlap{\;,}\label{eq:loop1}
\end{align}
where we used first \cref{eq:boundaryqB2} and then \cref{eq:boundaryqB1}. On the other hand,
\begin{align}
\MoveEqLeft\wavefunction(z+\tau+1,\tau,\zeta)=\exp\bigg(\frac{\I\,qB\im(z+\zeta+\tau)}{2\im\tau}\bigg)
 \wavefunction(z+\tau,\tau,\zeta)\nonumber\\
 &=\exp\bigg(\I\frac{qB\im(z+\zeta+\tau)}{2\im\tau}\bigg)\exp\bigg(\frac{\I\,qB}{2\im\tau}\im\bar{\tau}(z+\zeta)\bigg)
 \wavefunction(z,\tau,\zeta)\nonumber\\
 &=\exp\bigg(\I\frac{q B\im\tau }{2\im\tau}\bigg)\exp\bigg(\frac{\I\,qB}{2\im\tau}\im(\bar{\tau}(z+\zeta))\bigg)
 \exp\bigg(\frac{\I\,qB}{2\im\tau}\im(z+\zeta)\bigg)
 \wavefunction(z,\tau,\zeta)\;,\label{eq:loop2}
\end{align}
where we used first \cref{eq:boundaryqB1} and then \cref{eq:boundaryqB2}.
Imposing that \cref{eq:loop1} and \cref{eq:loop2} yield the same wave function
leads to the flux quantization condition
\begin{gather}
    qB = 2\pi M\;, \label{eq:qconditionM}
\end{gather}
with $M$ an arbitrary integer. Therefore, in what follows, we will not
consider $q$ and $B$ individually, but only the integer $M$ instead. Then, we have
\begin{subequations}
\begin{align}
    \wavefunction(z+1,\tau,\zeta)&=
	\exp\bigg(\frac{\I\,\pi M}{\im\tau}\im(z+\zeta)\bigg)\,\wavefunction(z,\tau,\zeta)
    \label{eq:boundary1}\;,\\
    \wavefunction(z+\tau,\tau,\zeta )&=
	\exp\bigg(\frac{\I\,\pi M}{\im\tau}\im(z+\zeta)\bar{\tau}\bigg)\,\wavefunction(z,\tau, \zeta )
	\;.
    \label{eq:boundary2}
\end{align}
\end{subequations}

\section{Yukawa couplings}
\label{sec:YukawaCouplings}

\subsection{Couplings from overlap integrals}

One of the main rationales of working out the wave functions in
\cref{sec:zero-modes} is that the overlaps of wave functions yield the (Yukawa)
couplings of the model. Let us consider a $4+2$ dimensional theory which is
compactified on a torus $\mathbb{T}^2$. There is a gauge group breaking
$\U{N}\rightarrow \U{N_{a}}\times \U{N_{b}}\times \U{N_{c}}$ with
$N=N_{a}+N_{b}+N_{c}$ due to the introduction of a magnetic flux in the compact
dimensions given by
\begin{equation}\label{eq:fluxF}
F_{z\bar{z}}=\frac{\pi \I}{\text{Im}\tau}\begin{pmatrix}
\frac{m_{a}}{N_{a}}\mathds{1}_{N_{a}\times N_{a}} & 0 & 0 \\
0 & \frac{m_{b}}{N_{b}}\mathds{1}_{N_{b}\times N_{b}} & 0 \\
0 & 0 & \frac{m_{c}}{N_{c}}\mathds{1}_{N_{c}\times N_{c}}
\end{pmatrix}\;,
\end{equation}
where we will assume that $s_{\alpha}=\frac{m_\alpha}{N_\alpha}$
is an integer for $\alpha\in\{a,b,c\}$. Then, in
\cite[equation~(5.7)]{Cremades:2004wa} one finds that Yukawa couplings of the
4D effective theory are given by
\begin{equation}\label{eq:Yijk}
 Y_{ijk}(\widetilde{\zeta},\tau)=g\,\sigma_{abc}\,\int\limits_{\mathds{T}^2}\!\D^2 z\,
 \wavefunction[i,\mathcal{I}_{ab}](z,\tau,\zeta_{ab})\,
 \wavefunction[j,\mathcal{I}_{ca}](z,\tau,\zeta_{ca})\,
 \left(\wavefunction[k,\mathcal{I}_{cb}](z,\tau,\zeta_{cb})\right)^*\;.
\end{equation}
Here, $\wavefunction[i,\mathcal{I}_{ab}](z,\tau,\zeta_{ab})$ are the wave
functions of \cref{eq:ZeroModesWavefunctions} that represent chiral fermions
bifundamentals transforming as $(\rep{N}_{a},\crep{N}_{b})$ under
$\U{N_a}\times\U{N_b}$, and similarly for $\wavefunction[j,\mathcal{I}_{ca}]$
and $\wavefunction[k,\mathcal{I}_{cb}]$. The multiplicities of
$\mathcal{I}_{\alpha\beta}$ of these bifundamentals are given by
\begin{equation}\label{eq:I_alpha_beta}
 \mathcal{I}_{\alpha\beta}=s_\alpha-s_\beta\;,
\end{equation}
which implies that
\begin{equation}\label{eq:FluxCondition}
\mathcal{I}_{ab}+\mathcal{I}_{bc}+\mathcal{I}_{ca}=0\;.
\end{equation}
Furthermore, $g$ is the $(4+2)$--dimensional gauge coupling, and
$\sigma_{abc}=\operatorname{sign}(\mathcal{I}_{ab}\mathcal{I}_{bc}\mathcal{I}_{ca})$
~\cite{Cremades:2004wa} is a sign which is equal to $-1$ throughout our
discussion. The $\zeta_{\alpha\beta}$ are given by
\begin{equation}
 \zeta_{\alpha\beta}=\frac{s_{\alpha}\zeta_{\alpha}-s_{\beta}\zeta_{\beta}}{s_{\alpha}-s_{\beta}}
\end{equation}
for $\alpha,\beta\in\{a,b,c\}$. Finally, $\zeta_{\alpha}$ are the Abelian Wilson
lines associated to the group $\U{N_{\alpha}}$ for $\alpha\in\{a,b,c\}$.
$\zeta_{cb}$ and $\zeta_{ca}$ are defined similarly. As one can see from
\cref{eq:ZeroModesWavefunctions}, the $\zeta_{\alpha}$ represent translation of
the torus origin. However, as shown in \cite{Cremades:2004wa} if all three wave
functions are shifted by the same Wilson line,  then the values of the Yukawa
couplings are unaffected.

\subsection{Yukawa couplings for generic flux parameters}

Let us now discuss how one can reduce the overlap integrals \eqref{eq:Yijk} to a
linear combination of $\vartheta$--functions. We follow the strategy of
\cite{Cremades:2004wa}, but generalize the result to the cases
$\mathcal{I}_{ab}>1$ and/or
$\gcd(\mathcal{I}_{ab},\mathcal{I}_{ca},\mathcal{I}_{bc})>1$, with
$\mathcal{I}_{ab},\mathcal{I}_{ca}>0$ and $\mathcal{I}_{bc}<0$. Note that
the analogous discussion applies to the case in which $\mathcal{I}_{ab}$ and
$\mathcal{I}_{ca}$ are negative \cite[cf.\ the discussion around
equation~(5.6)]{Cremades:2004wa}.

In order to find closed--form expressions for the Yukawa
couplings, one uses two important facts~\cite{Cremades:2004wa}:
\begin{enumerate}
 \item products of $\vartheta$--functions can be expanded in terms of
 $\vartheta$--functions, see~\cite[equation~(5.8)]{Cremades:2004wa}, and that 
 \item the $\vartheta$--functions fulfill certain orthogonality and completeness
 relations.
\end{enumerate} 
These facts allow one to find analytic expressions for the Yukawa couplings
\eqref{eq:Yijk} that do no longer involve integrals~\cite{Cremades:2004wa}. In
more detail, starting from \eqref{eq:Yijk}, one obtains
(cf.\ \cite[equation~(5.15)]{Cremades:2004wa}) 
\begin{align}\label{eq:CIM5.15a}
 Y_{ijk}(\widetilde{\zeta},\tau)&=\mathcal{N}_{abc}\,
 \mathrm{e}^{ \frac{H(\widetilde{\zeta},\tau)}{2}}\,
 \sum\limits_{m\in\mathds{Z}_{\mathcal{I}_{bc}}}
 \delta_{k,i+j+\mathcal{I}_{ab}\,m}\,
 \vartheta\Orb{\frac{\mathcal{I}_{ca}i-\mathcal{I}_{ab}j+\mathcal{I}_{ab}\mathcal{I}_{ca}m}{
	-\mathcal{I}_{ab}\mathcal{I}_{bc}\mathcal{I}_{ca}}}{0}
	\bigl(\widetilde{\zeta},\tau\,|\mathcal{I}_{ab}\mathcal{I}_{bc}\mathcal{I}_{ca}|\bigr)\;,
\end{align}
where
\begin{equation}
 \mathcal{N}_{abc}=g\,\sigma_{abc}\,\left(\frac{2\im\tau}{\mathcal{A}^2}\right)^{1/4}
 \,\left|\frac{\mathcal{I}_{ab}\mathcal{I}_{ca}}{\mathcal{I}_{bc}}\right|^{1/4}
\end{equation}
is a normalization constant and the Wilson line dependence is encoded in the quantities
\begin{align}
 \widetilde{\zeta}&:=-\mathcal{I}_{ab}\,\mathcal{I}_{ca}\,(\zeta_{ca}-\zeta_{ab})
 =d^{\alpha\beta\gamma}\,s_\alpha\,\zeta_\alpha\,\mathcal{I}_{\beta\gamma}
 \label{eq:zeta-tilde}
\end{align}
and
\begin{align}
 \frac{H(\widetilde{\zeta},\tau)}{2}
 &:=\frac{\pi\I}{\im \tau}\left(\mathcal{I}_{ab}\,\zeta_{ab}\,\im\zeta_{ab}
 +\mathcal{I}_{bc}\,\zeta_{bc}\,\im\zeta_{bc}
 +\mathcal{I}_{ca}\,\zeta_{ca}\,\im\zeta_{ca}\right)\notag\\
 &=\frac{\pi\I}{\im\tau}\left|\mathcal{I}_{ab}\,\mathcal{I}_{bc}\,\mathcal{I}_{ab}\right|^{-1}
 \frac{\widetilde{\zeta}\,\im\widetilde{\zeta}}{\im\tau}
 \;.\label{eq:WilsonH}
\end{align}
with
\begin{equation}
 d^{\alpha\beta\gamma}=
 \begin{cases}
  1\;,&\text{if $\{\alpha,\beta,\gamma\}$ is an even permutation of $\{1,2,3\}$}\;,\\
  0\;, &\text{otherwise}\;,
 \end{cases}
\end{equation}
where we have used~\cite[equation~(5.28)]{Cremades:2004wa}. Cremades et
al.\ obtain then \cite[equation~(5.15)]{Cremades:2004wa}
\begin{align}\label{eq:CIM5.15b}
 Y_{ijk}(\widetilde{\zeta},\tau)&=\mathcal{N}_{abc}\mathrm{e}^{ \frac{H(\widetilde{\zeta},\tau)}{2}}\,
 \vartheta\Orb{-\left(\frac{j}{\mathcal{I}_{ca}}+\frac{k}{\mathcal{I}_{bc}}\right)
 	/\mathcal{I}_{ab}}{0}
	\bigl(\widetilde{\zeta},\tau\,|\mathcal{I}_{ab}\mathcal{I}_{bc}\mathcal{I}_{ca}|\bigr)
	\qquad\text{for }
	i=k-j\mod \mathcal{I}_{ab}\;.
\end{align}
This expression yields the correct couplings only if $\mathcal{I}_{ab}=1$,
which implies that $d=1$, where
\begin{equation}\label{eq:d=gcd}
 d:=\gcd\bigl(|\mathcal{I}_{ab}|,|\mathcal{I}_{ca}|,|\mathcal{I}_{bc}|\bigr)\;. 
\end{equation} 
To see that we need to demand that $d=1$ for \eqref{eq:CIM5.15b} to hold, notice
that in \eqref{eq:CIM5.15a} the integers $i$, $j$ and $k$ are only defined
modulo $\mathcal{I}_{ab}$, $\mathcal{I}_{bc}$ and $\mathcal{I}_{ca}$,
respectively. This is evident from the overlap integral \eqref{eq:Yijk}, where
e.g.\ $\wavefunction[i,\mathcal{I}_{ab}](z,\tau,\zeta_{ab})=
\wavefunction[i+\mathcal{I}_{ab},\mathcal{I}_{ab}](z,\tau,\zeta_{ab})$. However,
if $\gcd(|\mathcal{I}_{ab}|,|\mathcal{I}_{ca}|)>1$ or $|\mathcal{I}_{ab}|>1$,
shifting $i$ (or $j$) by $|\mathcal{I}_{ab}|$ (or $(|\mathcal{I}_{ca}|$), which
leaves the wave functions invariant and hence has to produce the same overlap
integral, leads to different results for the Yukawa couplings when using
\eqref{eq:CIM5.15b}.

To obtain the general expression, let us look at
\cite[Proposition II.6.4.\ on p.\ 221]{Mumford:Theta1}
\begin{multline}
 \vartheta\orb{\frac{j}{\mathcal{I}_{ab}}}{0}(z_1,\mathcal{I}_{ab}\tau)\cdot
 \vartheta\orb{\frac{j}{\mathcal{I}_{ca}}}{0}(z_2,\mathcal{I}_{ca}\tau)
 =\sum\limits_{\mathclap{m\in\mathds{Z}_{\mathcal{I}_{ab}+\mathcal{I}_{ca}}}}
 \vartheta\Orb{\frac{i+j+\mathcal{I}_{ab}m}{\mathcal{I}_{ab}+\mathcal{I}_{ca}}}{0}\bigl(z_1+z_2,(\mathcal{I}_{ab}+\mathcal{I}_{ca})\tau\bigr)
 \\
 \vartheta\Orb{\frac{\mathcal{I}_{ca}i-\mathcal{I}_{ab}j+\mathcal{I}_{ab}\mathcal{I}_{ca}\,m}{\mathcal{I}_{ab}\mathcal{I}_{ca}(\mathcal{I}_{ab}+\mathcal{I}_{ca})}}{0}
 \left(\mathcal{I}_{ca}\,z_1-\mathcal{I}_{ab}\,z_2,\mathcal{I}_{ab}\mathcal{I}_{ca}(\mathcal{I}_{ab}+\mathcal{I}_{ca})\tau\right)\;,
 \label{eq:MumfordMultiplyingThetasUpper}
\end{multline}
which was used in \cite{Cremades:2004wa}. In our wave functions,
$z_1=\mathcal{I}_{ab}\,(z+\zeta_{ab})$ and $z_2=\mathcal{I}_{ca}\,(z +\zeta_{ca})$, so that in the overlap
integral $z_1+z_2=\mathcal{I}_{cb}\,(z+\zeta_{cb})$ and
$\mathcal{I}_{ca}\,z_1-\mathcal{I}_{ab}\,z_2=\widetilde{\zeta}$.  One thus
obtains (cf.\ \cite[equation~(5.12)]{Cremades:2004wa})
\begin{multline}
 \wavefunction[i,\mathcal{I}_{ab}](z,\tau,\zeta_{ab} )\cdot
 \wavefunction[j,\mathcal{I}_{ca}](z,\tau,\zeta_{ca} )
 =\mathcal{A}^{-1/2}\,(2\im\tau)^{1/4}\,
 \left|\frac{\mathcal{I}_{ab}\,\mathcal{I}_{ca}}{\mathcal{I}_{bc}}\right|^{1/4}\mathrm{e}^{ \frac{H(\widetilde{\zeta},\tau)}{2}}
 \\
 \sum\limits_{\mathclap{m\in\mathds{Z}_{|\mathcal{I}_{bc}|}}}
 \wavefunction[i+j+\mathcal{I}_{ab}\,m,\mathcal{I}_{cb}](z,\tau,\mathcal{I}_{cb})
 \,\vartheta\Orb{\frac{\mathcal{I}_{ca}i-\mathcal{I}_{ab}j+\mathcal{I}_{ab}\mathcal{I}_{ca}\,m}{\mathcal{I}_{ab}\mathcal{I}_{ca}(\mathcal{I}_{ab}+\mathcal{I}_{ca})}}{0}
 \left(\widetilde{\zeta},\mathcal{I}_{ab}\mathcal{I}_{ca}(\mathcal{I}_{ab}+\mathcal{I}_{ca})\tau\right)
 \;.\label{eq:CIM5.12}
\end{multline}
The product \eqref{eq:CIM5.12} gets projected on a third wave function
$\psi^{k,\mathcal{I}_{ab}+\mathcal{I}_{ca}}$ via the overlap integral
\eqref{eq:Yijk}. This means that $k$ has to ``match'', i.e.\ $m$ has to be
a solution of the congruence equation
\begin{equation}\label{eq:linearm}
    \mathcal{I}_{ab}\, m +i+j= k \mod \mathcal{I}_{bc}\;.
\end{equation}
Now observe that, since
$\mathcal{I}_{ab}+\mathcal{I}_{bc}+\mathcal{I}_{ca}=0$,
$\gcd\bigl(|\mathcal{I}_{ab}|,|\mathcal{I}_{cb}|\bigr)=d$ with $d$ from
\cref{eq:d=gcd}. \Cref{eq:linearm} is a linear congruence equation for the
variable $m$. It is known that  (cf.\ e.g.\ \cite[Lemma 3 on
p.~37]{Dudley:2012ent}) if
\begin{equation}\label{eq:ijk_selection_rule_1}
 k-i-j = 0 \mod d\,,
\end{equation}
the linear congruence of \cref{eq:linearm} has $d$ solutions. Otherwise there is
no solution. Note that the condition~\eqref{eq:ijk_selection_rule_1} provides us
with a selection rule for the Yukawa couplings, which can be interpreted as a
$\mathds{Z}_d$ flavor symmetry (cf.\ \cite{Abe:2009uz}). We thus know that
the Yukawa couplings will be proportional to
\begin{equation}\label{eq:modulo_Kronecker_delta}
 \Delta_{i+j,k}^{(d)}:=
 \begin{dcases}
  1\;,&\text{if }i+j=k\mod d\;,\\
  0\;,&\text{otherwise}\;.
 \end{dcases}
\end{equation}
Consider now combinations of $i$, $j$ and $k$ satisfying the selection rule
\eqref{eq:ijk_selection_rule_1}. This means that 
\begin{equation}
 k-i-j=m'\,d
\end{equation}
with some integer $m'=(k-i-j)/d$. Define now
$\mathcal{I}_{ab}'=\mathcal{I}_{ab}/d$,  $\mathcal{I}_{ca}'=\mathcal{I}_{ca}/d$ 
and $\mathcal{I}_{bc}'=\mathcal{I}_{bc}/d$, which are integers because of
\cref{eq:d=gcd}. We can thus divide \cref{eq:linearm} by $d$ to get
\begin{equation}
    |\mathcal{I}_{ab}'|\, m = m' \mod |\mathcal{I}_{bc}'|\;,
    \label{eq:linearm2}
\end{equation}
where $\gcd\bigl(|\mathcal{I}_{ab}'|,|\mathcal{I}_{bc}'|\bigr)=1$. 
\Cref{eq:linearm2} can be solved with e.g.\ the \texttt{Mathematica}
command \texttt{FindInstance}. However, as we shall discuss now, one can find a
closed--form expression for the solution.
The linear congruence \eqref{eq:linearm2} has one (inequivalent) solution
$m=m_{0}$, which is given by
$\bigl[|\mathcal{I}_{ab}'|\bigr]_{(|\mathcal{I}_{bc}'|)}m'$ where
$\bigl[|\mathcal{I}_{ab}'|\bigr]_{(|\mathcal{I}_{bc}'|)}$ is the  multiplicative
inverse of $|\mathcal{I}_{ab}'|$ modulo $|\mathcal{I}_{bc}'|$.  According to
Euler's theorem (cf.\ e.g.\ \cite[Theorem 1 on p.~64]{Dudley:2012ent}), the
multiplicative inverse can be expressed via the Euler $\phi$--function,
$\bigl[|\mathcal{I}_{ab}'|\bigr]_{(|\mathcal{I}_{bc}'|)}=(\mathcal{I}_{ab}')^{\phi\bigl(|\mathcal{I}_{bc}'|\bigr)-1}$.
This means that
\begin{equation}\label{eq:m0_via_Euler_1}
 m_0=(\mathcal{I}_{ab}')^{\phi\bigl(|\mathcal{I}_{bc}'|\bigr)-1}\,
 \frac{k-i-j}{d}\mod |\mathcal{I}_{bc}'|\;.
\end{equation}  
Note that the Euler $\phi$--function is implemented in \texttt{Mathematica} as
\texttt{EulerPhi}. Relation \eqref{eq:m0_via_Euler_1} implies that one
particular solution $m_0$ of \cref{eq:linearm} satisfies
\begin{equation}
 \mathcal{I}_{ab}\,m_0=
 (\mathcal{I}_{ab}')^{\phi\bigl(|\mathcal{I}_{bc}'|\bigr)}\,
 (k-i-j)\mod |\mathcal{I}_{bc}|\;.
\end{equation}
Given the solution $m_0$ in \cref{eq:m0_via_Euler_1}, the $d$ solutions of
\cref{eq:linearm} are given by 
\begin{equation}\label{eq:solutionsm}
 m=m_{0}-|\mathcal{I}_{bc}'|\,t\quad\text{for }t=0,\dots, (d -1)\;. 
\end{equation}
Thus, using \cref{eq:solutionsm} in \eqref{eq:CIM5.15a}, we see that the 
Yukawa couplings are given by
\begin{align}
    Y_{ijk}(\widetilde{\zeta},\tau) &=\mathcal{N}_{abc}\mathrm{e}^{ \frac{H(\widetilde{\zeta},\tau)}{2}} \,\Delta_{i+j,k}^{(d)}
	\,\sum_{t=0}^{d-1}
	\vartheta\Orb{\frac{\mathcal{I}_{ca}i-\mathcal{I}_{ab}j+\mathcal{I}_{ab}\mathcal{I}_{ca}m_{0}}{|\mathcal{I}_{ab} \mathcal{I}_{bc}\mathcal{I}_{ca}|}+\frac{t}{d}}{0}
	\left(\widetilde{\zeta},|\mathcal{I}_{ab}\mathcal{I}_{ca}\mathcal{I}_{bc}|\tau
	\right)
	\;,\label{eq:Yijk_2}
\end{align}
\Cref{eq:Yijk_2} can be simplified further. Let us define
\begin{subequations}
\begin{align}
 P&:=|\mathcal{I}_{ab}\mathcal{I}_{ca}\mathcal{I}_{bc}|\;,\\
 \lambda&:=\lcm\bigl(|\mathcal{I}_{ab}|,|\mathcal{I}_{ca}|,|\mathcal{I}_{bc}|\bigr)\;.
 \label{eq:def_L_lcm}
\end{align}
\end{subequations}
Next we note that\footnote{To see this, consider two positive integers $a$ and $b$, and define
$c=\gcd(a,b)=\gcd\bigl(a,b,(a+b)\bigr)$ such that $a=a'\,c$ and $b=b'\,c$ with
integers $a'$ and $b'$. Then $\lcm\bigl(a,b,(a+b)\bigr)=c\,
\lcm\bigl(a',b',(a'+b')\bigr)$. Since $a'$, $b'$ and $(a'+b')$ do not have a
nontrivial common divisor, 
\[ \lcm\bigl(a,b,(a+b)\bigr)=c\,a'\,b'\,(a'+b')\;,\]
so that
\[
 a\,b\,(a+b)=[\gcd\bigl(a,b,(a+b)\bigr)]^2\lcm\bigl(a,b,(a+b)\bigr)\;.
\]}
\begin{equation}\label{eq:P=lambdad2}
 P= \lambda\,d^2\;.
\end{equation}
Then \cref{eq:Yijk_2} can be recast as 
\begin{equation}\label{eq:sumForYukawas}
    Y_{ijk}(\widetilde{\zeta},\tau) 
	=\mathcal{N}_{abc}\mathrm{e}^{ \frac{H(\widetilde{\zeta},\tau)}{2}}\, 
	\sum_{t=0}^{d-1}
	\vartheta\Orb{\frac{1}{d}\left(\frac{\widehat{\alpha}_{ijk}}{\lambda}+t\right)}{0}
	\left(\widetilde{\zeta},P\tau\right)\;,
\end{equation}
where
\begin{equation}
\widehat{\alpha}_{ijk}=\mathcal{I}_{ca}'\,i-\mathcal{I}_{ab}'\,j
+\mathcal{I}_{ca}'\,\mathcal{I}_{ab}\,m_0
\end{equation}
is an integer. Using \cref{eq:m0_via_Euler_1}, $\widehat{\alpha}_{ijk}$ becomes
\begin{equation}
\widehat{\alpha}_{ijk}=\mathcal{I}_{ca}'\,i
-\mathcal{I}_{ab}'\,j+
\mathcal{I}_{ca}'\,(\mathcal{I}_{ab}')^{\phi\bigl(|\mathcal{I}_{bc}'|\bigr)}\,
( k-i-j )\mod \lambda\,d
\;.
\end{equation}
Now we can use \cref{eq:ThetaSum1} to express the sum~\eqref{eq:sumForYukawas}
as
\begin{align}
    Y_{ijk}(\widetilde{\zeta},\tau) 
	&=\mathcal{N}_{abc}\mathrm{e}^{ \frac{H(\widetilde{\zeta},\tau)}{2}}\, 
	\sum_{t=0}^{d-1}
	\sum_{\ell=-\infty}^{\infty}
	\exp\left[\I\pi\left(\frac{1}{d}\frac{\widehat{\alpha}_{ijk}}{\lambda}+\frac{1}{d}t+\ell\right)^2P\,\tau\right] \,
	\exp\left[2\pi\I\,\left(\frac{\widehat{\alpha}_{ijk}}{\lambda\,d}+\frac{t}{d}+\ell \right)\,\widetilde{\zeta}\right]
	\notag\\
	&=\mathcal{N}_{abc}\mathrm{e}^{ \frac{H(\widetilde{\zeta},\tau)}{2}}\, 
	\sum_{\ell=-\infty}^{\infty}
	\sum_{t=0}^{d-1}
	\exp\left[\I\pi\left(\frac{\widehat{\alpha}_{ijk}}{\lambda}+t
	+d\,\ell\right)^2
	\lambda\,\tau\right]
	\,\exp\left[2\pi\I\,\left(\frac{\widehat{\alpha}_{ijk}}{\lambda}+t
	+\ell d \right)\,\frac{\widetilde{\zeta}}{d}\right]
	\notag\\
		&=\mathcal{N}_{abc}\mathrm{e}^{\frac{H(\widetilde{\zeta},\tau)}{2}}\, 
	\sum_{\ell'=-\infty}^{\infty}
	\exp\left[\I\pi\left(\frac{\widehat{\alpha}_{ijk}}{\lambda}
	+\ell'\right)^2\lambda\,\tau\right]  
	\,\exp\left[2\pi\I\,\left(\frac{\widehat{\alpha}_{ijk}}{ \lambda\,d}+\ell' \right)\,\frac{\widetilde{\zeta}}{d}\right]
	\notag\\
		&=\mathcal{N}_{abc}\mathrm{e}^{ \frac{H(\widetilde{\zeta},\tau)}{2}}\,
	\vartheta\Orb{\frac{\widehat{\alpha}_{ijk}}{\lambda}}{0}
	\left(\frac{\widetilde{\zeta}}{d}, \lambda\,\tau\right)\;.
	\label{eq:theta_trickery_1}
\end{align}
Here, $\ell'=d\,\ell+t$. When $\ell$ runs over all integers, and $t$ runs from
$0$ to $d-1$, $\ell'$ runs over all integers. The $\widehat{\alpha}_{ijk}$ are
integers. Therefore, the physical Yukawa couplings are given by
\begin{align}\label{eq:explicit_Yukawa}
 Y_{ijk}(\widetilde{\zeta},\tau) &= \mathcal{N}_{abc}\,
 \mathrm{e}^{ \frac{H(\widetilde{\zeta},\tau)}{2}}
 \,\Delta_{i+j,k}^{(d)}
 \,\vartheta 
 \Orb{\frac{\mathcal{I}_{ca}'\,i-\mathcal{I}_{ab}'\,j
 +\mathcal{I}_{ca}'\,\left(\mathcal{I}_{ab}'\right)^{\phi\left(|\mathcal{I}_{bc}'|\right)}\,
  (k-i-j)}{\lambda}}{0}\left(\frac{\widetilde{\zeta}}{d}, \lambda\,\tau\right)
\end{align}
with $d$ from \cref{eq:d=gcd}, $\Delta_{i+j,k}^{(d)}$ from
\cref{eq:modulo_Kronecker_delta}, $\lambda$ from \cref{eq:def_L_lcm} and 
assuming $\mathcal{I}_{ab},\,\mathcal{I}_{ca} >0$ and $\mathcal{I}_{bc} <
0$.  Note that if $d=1$ and $\mathcal{I}_{ab}=1$, this formula reproduces
\cref{eq:CIM5.15b}. Further, \emph{a priori} this expression does not rely on
supersymmetry, it is simply derived from the overlap of wave functions. However,
one may expect the scalar wave function to be subject to substantial corrections
in non--supersymmetric theories. In \cref{sec:CommentsOnSUSY} we will argue that
magnetized tori may not comply with these expectations, and that this formula
may even be a good leading--order result in a non--supersymmetric theory. The
normalization factors in \cref{eq:explicit_Yukawa} are
\begin{equation}\label{eq:YukawaNormalization}
 \mathcal{N}_{abc}=g\,\sigma_{abc}\,
 \bigg(\frac{2\im\tau}{\mathcal{A}^{2}}\bigg)^{\nicefrac14}
		\, \lambda^{\nicefrac14}\,\left|\frac{1}{\mathcal{I}_{bc}'}\right|^{\nicefrac12}
\end{equation}
with $g$ being the gauge coupling. In
\cref{eq:RelationPhysicalVsHolomorphicYukawa} we will express the normalization
in terms of K\"ahler potential terms. Notice that if there are nontrivial
relative Wilson lines, the normalization of the fields changes compared to the
case without Wilson lines~\cite[equation~(7.37)]{Cremades:2004wa}. This has to
be taken into account when computing physical Yukawa couplings. In what follows,
we will set the Wilson lines to zero, leaving the detailed study of their impact
on the modular flavor symmetries for future work. As mentioned above, the
selection rule~\eqref{eq:ijk_selection_rule_1} entails a $\Z{d}$ symmetry. As we
discuss in more detail in \cref{app:IndependentYukawas}, out of \emph{a priori} 
$P=\lambda\,d^2$ entries, at most $\nicefrac{\lambda}{2} +1$ are distinct.

\section{Modular transformations}
\label{sec:ModularTransformations}

\subsection{Modular groups and modular forms}
\label{subsec:ModularGroupsForms}

The modular group $\Gamma=\SL{2,\mathds{Z}}$ can be defined by the 
presentation relations
\begin{equation}
  S^4=(S\,T)^3=\mathds{1}\quad\text{and}\quad S^2\,T = T\,S^2\;,
\end{equation}
where the generators $S$ and $T$ are usually chosen as
\begin{equation}\label{eq:SL2Z_S_T}
 S=\begin{pmatrix} 0 & 1 \\ -1 &0 \end{pmatrix}
 \quad\text{and}\quad
 T=\begin{pmatrix} 1 & 1 \\ 0 &1 \end{pmatrix}\;.
\end{equation}
These generators act on the modulus $\tau$ according to
\begin{equation}
 \tau\xmapsto{~S~}-\frac{1}{\tau}
 \quad\text{and}\quad
 \tau\xmapsto{~T~}\tau+1\;.
 \label{eq:Ttransformation}
\end{equation}
Hence, a general modular transformation acts on the modulus $\tau$ as
\begin{equation}
  \tau~\xmapsto{~\gamma~}~\frac{a\,\tau+b}{c\,\tau+d}~=:~\gamma\,\tau\;,\quad\text{where}\quad
  \gamma~=\begin{pmatrix} a & b \\ c & d \end{pmatrix}\in\Gamma\;,
  \label{eq:ModularTrafoTau}
\end{equation}
such that $ad-bc=1$ and $a,b,c,d\in\Z{}$. Consequently, functions of $\tau$ also
transform under $\gamma$. This is particularly true for modular forms, which are holomorphic
functions of $\tau$ (also at $\tau\to\I\infty$) with $\im\tau>0$~\cite{Bruinier:2008xxx}.
Modular forms $f_{\widehat\alpha}(\tau)$ of modular weight $k\in\mathds N$ and level $N=2,3,4,\ldots$
build finite vector spaces and transform under a modular transformation $\gamma\in\Gamma$ as~\cite{Liu:2019khw}
\begin{equation}
  f_{\widehat\alpha}(\tau)~\xmapsto{~\gamma~}~f_{\widehat\alpha}(\gamma\,\tau)~
     :=~(c\,\tau+d)^k\,\rho_{\boldsymbol r}(\gamma)_{\widehat\alpha\widehat\beta}\,f_{\widehat\beta}(\tau)\,,
  \label{eq:ModularTrafof}
\end{equation}
where $\widehat\alpha,\widehat\beta$ are considered here just as (integer) counters,
$(c\,\tau+d)^k$ often gets referred to as automorphy factor, and $\rho_{\boldsymbol r}(\gamma)$ 
denotes an $r$-dimensional (irreducible) representation matrix of $\gamma$ under the finite
modular group $\Gamma'_N\cong\SL{2,\Z{N}}$. These finite groups are defined
by the relations
\begin{equation}
   S^4 = (S\,T)^3 = \Id\,,\quad S^2\,T = T\,S^2\,,\quad T^N=\Id
\end{equation}
and an additional relation that ensures finiteness for $N>5$. 

It has been proposed in~\cite{Feruglio:2017spp} that Yukawa couplings in
quantum field theories can be modular forms, whereas, despite not being modular
forms, ``matter'' superfields  $\phi^i$ transform under a general modular
transformation $\gamma\in\Gamma$ as
\begin{equation}
  \phi^i~\xmapsto{~\gamma~}~(c\,\tau+d)^{k_\phi}\,\rho_{\boldsymbol s}(\gamma)_{ij}\,\phi^j\;.
  \label{eq:ModularTrafoMatter}
\end{equation}
Here $\rho_{\boldsymbol s}(\gamma)$ is the $s$-dimensional (reducible or
irreducible) $\Gamma'_N$  representation matrix. As for modular forms, the
powers $k_\phi$ are also known as modular weights  and are identical for the fields
in the transformation. Thus, matter fields build a  representation of the
finite modular group $\Gamma'_N$, which can be adopted as a symmetry of the
underlying (quantum) field theory. In this scenario, $\Gamma'_N$ can be
considered a ``modular flavor symmetry''.

In string--derived models, it is known that matter fields are subject to modular
transformations similar to \cref{eq:ModularTrafoMatter}. Moreover, Yukawa
couplings also transform as in \cref{eq:ModularTrafof}. However, as we shall see
in this section, the modular weights can be fractional and, hence, the emerging
modular flavor symmetry is not necessarily one of the $\Gamma'_N$. Yet, to
obtain fractional modular weights it is not necessary to go all the way to
strings, they already emerge from simpler settings such as magnetized tori (see
e.g.~\cite{Kobayashi:2016ovu,Kobayashi:2018rad,Kobayashi:2018bff,Kariyazono:2019ehj,Ohki:2020bpo,Kikuchi:2020frp,Kikuchi:2020nxn,Kikuchi:2021ogn}).
As we discuss in detail in detail in \cref{sec:ModularWeights}, this follows
already from the $\tau$--dependence of the normalization of the wave
functions~\cite{Cremades:2004wa}. 

As a first step, let us review the modular symmetries associated with modular
forms  with half--integral modular weights~\cite{Liu:2020msy}. In this case, one
must consider  instead of $\SL{2,\Z{}}$ its double cover, the so--called
metaplectic group $\widetilde{\Gamma}=\mathrm{Mp}(2,\mathds{Z})$. The generators
$\widetilde{S}$ and $\widetilde{T}$ of $\widetilde{\Gamma}$ satisfy the
presentation
\begin{equation}
\label{eq:Mp2Z-presentation}
  \widetilde{S}^8 = (\widetilde{S}\,\widetilde{T})^3=\mathds{1}
  \quad\text{and}\quad 
  \widetilde{S}^2\widetilde{T} = \widetilde{T}\,\widetilde{S}^2\;,
\end{equation}
which are represented by the choice
\begin{equation}
\label{eq:metaplecticGenerators}
   \widetilde{S} = (S,-\sqrt{-\tau})\quad\text{and}\quad 
   \widetilde{T} = (T,+1)\;,\qquad S,T\in\Gamma\;.
\end{equation}
In terms of these, the elements of the metaplectic group are given by
\begin{equation}\label{eq:Metaplecticgroup}
   \widetilde{\Gamma} = \left\{ \widetilde{\gamma} = (\gamma,\varphi(\gamma,\tau)) ~|~ 
     \gamma \in \Gamma,\ \varphi(\gamma,\tau) = \pm(c\,\tau+ d)^{\nicefrac12} \right\} \;,
\end{equation}
subject to the multiplication rule
\begin{equation}
\label{eq:metaplecticMultiplicationRule}
    \left(\gamma_1,\,\varphi(\gamma_1,\tau)\right)\left(\gamma_2,\,\varphi(\gamma_2,\tau)\right) 
    = \left(\gamma_1\gamma_2,\,\varphi(\gamma_1,\gamma_2\,\tau)\varphi(\gamma_2,\tau)\right)\;.
\end{equation}
To determine the sign of $\varphi(\gamma,\tau)$ for an arbitrary element
$\widetilde\gamma\in\widetilde\Gamma$, one has to express
$\widetilde\gamma$ as a product of the metaplectic
generators~\eqref{eq:metaplecticGenerators} and then use the multiplication
rule~\eqref{eq:metaplecticMultiplicationRule}. 

The modular transformations $\widetilde\gamma$ act on the modulus still just as
$\gamma$,  according to \cref{eq:ModularTrafoTau}. In contrast, modular forms of
modular weight  $\nicefrac{k}{2}$ and level $4N$, where $k,N\in\mathds{N}$,
transform as
\begin{equation}
  f_{\widehat\alpha}(\tau)~\xmapsto{~\widetilde\gamma~}~f_{\widehat\alpha}(\widetilde\gamma\,\tau)~:=~
     \varphi(\gamma,\tau)^k\,\rho_{\boldsymbol r}(\widetilde\gamma)_{\widehat\alpha\widehat\beta}\,f_{\widehat\beta}(\tau)
	 \;.
  \label{eq:MetaplecticTrafof}
\end{equation}
Here $\varphi(\gamma,\tau)^k$ is now the automorphy factor, and
$\rho_{\boldsymbol r}(\widetilde\gamma)$ is an (irreducible) representation
matrix of $\widetilde\gamma$ in the finite metaplectic modular group
$\widetilde{\Gamma}_{4N}$. The generators $\widetilde{S}$ and $\widetilde{T}$ of
this discrete group satisfy 
\begin{equation}
\label{eq:tildeGamma_4N-presentation}
  \widetilde{S}^8 = (\widetilde{S}\,\widetilde{T})^3=\Id\;,\quad
  \widetilde{S}^2\widetilde{T} = \widetilde{T}\,\widetilde{S}^2\;,\quad
  \widetilde{T}^{4N} = \Id
\end{equation}
and, for $N>1$, a relation to ensure the finiteness of the group. This amounts
to finding appropriate combinations of $\widetilde{S}$ and $\widetilde{T}$ that
yield $\Id_2\mod 4N$, where the modulo condition is to be understood
componentwise, and then demand that this combination yields identity in the
finite group. For $N=2$ we adopt the choice by~\cite[equation~(21)]{Liu:2020msy}
\begin{equation}
\label{eq:tildeGamma_4(2)-presentation}
   \widetilde{S}^5 \widetilde{T}^6 \widetilde{S}\widetilde{T}^4\widetilde{S}\widetilde{T}^2 \widetilde{S}\widetilde{T}^4 = \Id
   \;,
\end{equation}
and for $N=3$ we choose
\begin{equation}
\label{eq:tildeGamma_4(3)-presentation}
   \widetilde{S} \widetilde{T}^3 \widetilde{S}\widetilde{T}^{-2}\widetilde{S}^{-1}\widetilde{T} \widetilde{S}\widetilde{T}^{-3}
   \widetilde{S}^{-1}\widetilde{T}^2 \widetilde{S}^{-1}\widetilde{T}^{-1}= \Id\;.
\end{equation}
Note that $\widetilde\Gamma_{4N}$ is the double cover of $\Gamma'_{4N}$.
It is known that $\widetilde\Gamma_4\cong[96,67]$, $\widetilde\Gamma_8\cong[768, 1085324]$ 
and $\widetilde\Gamma_{12}$ is a group of order 2304. We use here the unique identifiers assigned
by the computer program \texttt{GAP}~\cite{GAP4}. For example, $[96,67]$ denotes a finite discrete
group of order 96, where 67 labels the group.

Finally, in field theories endowed with $\widetilde\Gamma_{4N}$ symmetries, 
the modular transformations of matter fields are given by
\begin{equation}
  \phi^i~\xmapsto{~\widetilde\gamma~}~\varphi(\gamma,\tau)^{k_\phi}\,\rho_{\boldsymbol s}(\widetilde\gamma)_{ij}\,\phi^j\;,
  \label{eq:MetaplecticTrafoMatter}
\end{equation}
where $\rho_{\boldsymbol s}(\widetilde\gamma)$ is now a (reducible or
irreducible) $\widetilde\Gamma_{4N}$  representation. As we shall see, this
behavior is natural in toroidal compactifications with magnetic fluxes.

\subsection{Normalization of the wave functions and modular weights}
\label{sec:ModularWeights}

The wave functions in \cref{eq:ZeroModesWavefunctions} satisfy
\begin{equation}\label{eq:NormalizationCondition}
 \int\limits_{\mathds{T}^2}\!\D^2z\,|\wavefunction(z,\tau,\zeta)|^2
 =
\mathcal{A}\,\int\limits_{0}^{1}\!\D x\,
 \int\limits_{0}^{1}\!\D y\,|\wavefunction(x+\tau\,y,\tau,\zeta)|^2
 \overset{!}{=}1\;,
\end{equation}
where $\mathds{T}^2$ denotes the fundamental domain of the torus, cf.\
\cref{sec:Normalization}. The normalization constant $\mathcal{N}\propto(\im\tau)^{-\nicefrac14}$ 
in \cref{eq:Normalization} is chosen in such a way that the normalization condition
\eqref{eq:NormalizationCondition} holds. This implies, in particular, that the
K\"ahler metric is proportional to $(\im\tau)^{-\nicefrac12}$, i.e.\
\begin{equation}\label{eq:MatterFieldKaehlerMetric}
 K_{i\bar\imath}\propto\frac{1}{\left(\im\tau\right)^{\nicefrac12}}\;,
\end{equation}
i.e.\ the modular weight of the 4D fields $\phi^{j,M}$ describing the zero modes
is $k_\phi=-\nicefrac{1}{2}$. We survey the  modular weights of the fields,
coupling and superpotential in \cref{tab:modweights}.
\begin{table}[t!]
	\centering
	\begin{tabular}{cccccc}
		\toprule
		object & $\wavefunction$ & $\phi^{j,M}$ & $\Omega^{j,M}$ & $Y_{ijk}$ & $\mathscr{W}$ \\
		\midrule 
		modular weight $k$ & $\nicefrac{1}{2}$ &  $-\nicefrac{1}{2}$ & $0$ & $\nicefrac{1}{2}$ & $-1$ \\
		\bottomrule
	\end{tabular}
	\caption{Modular weights of the $\mathds{T}^2$ wave functions
	$\wavefunction$, 4D fields $\phi^{j,M}$, 6D fields $\Omega^{j,M}$, Yukawa
	couplings $Y_{ijk}$, and superpotential $\mathscr{W}$.}
	\label{tab:modweights}
\end{table}
The modular weights $k_\psi$ of the wave functions can be
inferred from their normalization factor $\mathcal{N}$ in
\cref{eq:Normalization} to be $k_\psi=+\nicefrac{1}{2}$, as we shall also confirm
through their explicit modular transformations, \cref{eq:WaveFunctionTrafoSummary}. 
Therefore, the 6D fields,
\begin{equation}
    \Omega^{j,M} = \phi^{j,M}(x^{\mu}) \otimes \wavefunction(z,\tau)\;,
\end{equation}
have trivial modular weights, as they should. The modular weights of the Yukawa
couplings, $k_Y=\nicefrac{1}{2}$, can be explicitly determined from their
modular transformations, \cref{eq:T_transformation_Yukawa_theta,eq:S_transformation_Yukawa_theta}.
Since the superpotential terms
describing the Yukawa couplings involve three 4D fields and one coupling
``constant'', the superpotential $\mathscr{W}$ has modular weight
$k_{\mathscr{W}}=3k_\phi+k_Y=-1$. This means that under a modular transformation
the superpotential picks up an automorphy factor
\begin{equation}\label{eq:AutomorphyW}
 \mathscr{W}\xmapsto{~\gamma~}(c\,\tau+d)^{-1}\mathscr{W}\;.
\end{equation}
The automorphy factor $(c\,\tau+d)^{-1}$ can in general be ``undone'' by
so--called K\"ahler transformations~\cite{Wess:1992cp}, under which
\begin{subequations}\label{eq:KaehlerTrafo}
\begin{align}
 \mathscr{W}&\mapsto\mathrm{e}^{-\mathscr{F}(\Phi)}\,\mathscr{W}(\Phi)\;,\\
 K(\Phi,\overline{\Phi})&\mapsto
 K(\Phi,\overline{\Phi})+\mathscr{F}(\Phi)+\overline{\mathscr{F}(\Phi)}\;,
\end{align}
\end{subequations}
where $\Phi$ denotes the collection of 4D superfields, and $\mathscr F$ a
holomorphic function. In our case, the K\"ahler potential is, after setting the 
``matter'' fields to zero and at the classical level, given by (cf.\ e.g.\
\cite[equation~(5.50)]{Cremades:2004wa})
\begin{equation}\label{eq:KModuli}
 \widehat{K}=-\ln(\mathcal{S}+\overline{\mathcal{S}})
 -\ln(\mathcal{T}+\overline{\mathcal{T}})
 -\ln(\mathcal{U}+\overline{\mathcal{U}}) \subset K\;,
\end{equation}
in terms of the axio--dilaton $\mathcal{S}$, the K\"ahler modulus $\mathcal{T}$
and the complex structure modulus $\mathcal{U}$. These chiral fields are related to the
gauge coupling $g$, the torus volume $\mathcal{A}$ and $\tau$ according to
$\re{\mathcal S}\propto 1/g^2$, $\re{\mathcal T}\propto \mathcal{A}$ and $\re
\mathcal{U}=\im\tau$. Consequently, $\tau$ appears in the K\"ahler potential as
\begin{equation}\label{eq:ClassicalKaehlerPotentialTau}
 -\ln(\mathcal{U}+\overline{\mathcal{U}})=-\ln(-\I\,\tau+\I\,\bar\tau)\;.
\end{equation} 
Given that
\begin{equation}\label{eq:ModularTrafoTau-TauBar}
 \tau-\bar\tau\xmapsto{~\gamma~}|c\,\tau+d|^{-2}\,(\tau-\bar\tau)\;,
\end{equation}
it is easy to see that $K$ under a modular transformation of $\tau$ becomes
\begin{equation}
\label{eq:modTrafoK}
   K \xmapsto{~\gamma~} K + \ln(c\,\tau+d) + \ln(c\,\bar\tau+d)\;.
\end{equation}
A K\"ahler transformation~\eqref{eq:KaehlerTrafo} with $\mathscr{F} =
-\ln(c\,\tau+d)$ then absorbs simultaneously the modular transformation of $K$
and $\mathscr W$, see \cref{eq:AutomorphyW}, yielding a modular invariant
supersymmetric theory. That is, the supergravity K\"ahler function 
\begin{equation}
 G(\Phi,\overline{\Phi})=K(\Phi,\overline{\Phi})
 +\ln\left|\mathscr{W}(\Phi)\right|^2
\end{equation}
is automatically invariant under the simultaneous transformation
\eqref{eq:AutomorphyW} and \eqref{eq:modTrafoK}. In other words, we cannot dial
the modular weight of the superpotential at will, it is already determined by
the (classical) K\"ahler potential of the torus~\eqref{eq:KModuli}.  In
particular, setting the modular weight of the superpotential to zero is  not an
option in this approach, in which we derive modular flavor symmetries  from an
explicit torus.

\subsection{Boundary conditions for transformed wave functions}
\label{sec:mfs-even-odd}

It has been stated in the literature \cite{Ohki:2020bpo,Kikuchi:2020frp} that
the wave functions given by \cref{eq:ZeroModesWavefunctions} do not satisfy the
boundary conditions given by the lattice periodicity when transformed under
\cref{eq:Ttransformation} for odd units of flux, $M$. If true, this would mean
that a physical wave function gets mapped to an unphysical one just by looking at
an equivalent torus, which would indicate that either the expressions for the
wave functions were incorrect, or there is something fundamentally wrong with odd
$M$. In this case, simple explanations of three generations would be at stake.

However, as we shall see, the transformed wave functions do obey the correct
boundary conditions, both for even and odd $M$. The important point is that, if
our original wave function $\wavefunction(z,\tau,0)$ satisfied conditions for
$\tau$, after a modular transformation $\tau\mapsto\tau'$ the transformed
wave function $\wavefunction (z,\tau',0)$ needs to fulfill the
conditions for $\tau'$, and not for $\tau$.

For the modular $S$ transformations, the boundary conditions, given by
\cref{eq:boundary1,eq:boundary2}, are now
\begin{subequations}
\begin{align}
\wavefunction\left(-\frac{z}{\tau}+1,-\frac{1}{\tau} , 0\right)&=
 \exp\bigg(\I \pi M \frac{\im (-\nicefrac{z}{\tau})}{\im (-\nicefrac{1}{\tau})}   \bigg)\, 
\wavefunction \bigg(-\frac{z}{\tau},-\frac{1}{\tau} , 0 \bigg)\nonumber\\
  &= \exp\bigg(-\I \pi M \frac{\im z \bar{\tau}}{\im \tau}   \bigg)\, 
 \wavefunction \bigg(-\frac{z}{\tau},-\frac{1}{\tau} , 0 \bigg)\;,
    \label{eq:boundary1STransformed} \\
    \wavefunction \bigg(-\frac{z}{\tau}-\frac{1}{\tau},-\frac{1}{\tau} , 0\bigg)&=
 \exp\bigg(\I \pi M \frac{\im (-\nicefrac{z}{\tau}) (-\nicefrac{1}{\bar{\tau}})}{\im (-\nicefrac{1}{\tau})}   \bigg) \, \wavefunction \bigg(-\frac{z}{\tau},-\frac{1}{\tau}, 0 \bigg)\nonumber\\
 &= \exp\bigg(\frac{\I \pi M \im z }{\im \tau}\bigg)\,
 \wavefunction \bigg(-\frac{z}{\tau},-\frac{1}{\tau}, 0\bigg)\; .\label{eq:boundary2STransformed}
\end{align}
\end{subequations}
The fact that the transformed wave functions follow the boundary condition is a
consequence of the fact that the wave functions are functions of $z$ and $\tau$,
which we can just replace by their image under $S$. Nonetheless we verify this
explicitly in \cref{app:boundary_conditions_explicit_check_S}.

Next, under the modular $T$ transformation given by \cref{eq:Ttransformation}
the transformed boundary conditions, \cref{eq:boundary1,eq:boundary2}, are
\begin{subequations}
\begin{align}
\wavefunction (z+1,\tau+1,0)&=
 \exp\bigg(\I\,\frac{\pi M}{\im\tau}{\im z}\bigg)\, 
\wavefunction (z,\tau + 1,0)\;,
     \label{eq:boundary1TTransformed}\\
   \wavefunction (z+\tau+1,\tau+1,0)&=
 \exp\bigg(\I\,\frac{\pi M}{\im\tau}\im\bigl((\bar{\tau}+1)z\bigr)\bigg)\, 
\wavefunction (z,\tau +1,0)
	\;.
   \label{eq:boundary2TTransformed}
\end{align}
\end{subequations}
We can make the same argument as above but also verify the statement explicitly
in \cref{app:boundary_conditions_explicit_check_T}.

However, the transformed wave function \cref{eq:WavefunctionTtransformed}, i.e.\
the wave functions ``living'' on a torus with torus parameter $\tau'=\tau+1$  do
not follow the original boundary conditions of  \cref{eq:boundary1,eq:boundary2}
with $\tau$. Indeed, from \cref{eq:WavefunctionTtransformed} we get
\begin{align}
\wavefunction (z+\tau,\tau +1,0)&=\widetilde{\mathcal{N}}\mathrm{e}^{\frac{\I\,\pi M}{\im\tau}[z\im z+z\im\tau+\tau\im z+\tau\im\tau ]}
	\vartheta\orb{\frac{j}{M}}{\frac{M}{2}}(Mz+M\tau,M\tau)\nonumber\\
    &=\widetilde{\mathcal{N}} \mathrm{e}^{\frac{\I\,\pi M}{\im\tau}[z\im z+z\im\tau+\tau\im z+\tau\im\tau ]}\mathrm{e}^{-\I\,\pi M\tau-2\pi\I\,(Mz+\frac{M}{2})}
	\vartheta\orb{\frac{j}{M}}{\frac{M}{2}}(Mz,M\tau)\nonumber\\
    &=\mathrm{e}^{-\pi\I\,M}\mathrm{e}^{\frac{\I\,\pi M}{\im\tau}(z\im\tau+\tau\im z+\tau\im\tau-\tau\im\tau-2z\im\tau)}\widetilde{\mathcal{N}} \mathrm{e}^{i\pi M z \frac{\im z}{\im\tau}}
	\vartheta\orb{\frac{j}{M}}{\frac{M}{2}}(Mz,M\tau ) \nonumber\\
  &=\mathrm{e}^{-\pi\I\,M}\,\exp\bigg(\I\,\frac{M\pi}{\im\tau}\im\bar{\tau}z\bigg)
	\,\wavefunction (z,\tau +1,0)
	\;,
\end{align}
where 
$\widetilde{\mathcal{N}}:=\mathrm{e}^{-\I\pi\,j\bigl(1-\frac{j}{M}\bigr)}\mathcal{N}$ 
and we have used \cref{eq:z+ntau} in the second line. Thus, we find that
\begin{equation}
   \wavefunction (z+\tau,\tau +1,0)=\mathrm{e}^{-\pi\I\,M}
	\exp\bigg(\I\,\frac{M\pi}{\im\tau}\im\bar{\tau}z\bigg)\,
	\wavefunction (z,\tau + 1,0)\;.
    \label{eq:wrongboundary}
\end{equation}
Therefore, for odd $M$ \cref{eq:wrongboundary} differs from \cref{eq:boundary2}
by a phase. However, there is also no reason why the transformed wave functions
should obey boundary conditions for $\tau$ instead of $\tau'=\tau+1$.
Nevertheless, this fact will have important implications for the explicit form
of the $T$--transformation, as we shall see in
\cref{sec:ModularTransformationsOfWaveFunctions}.

\subsection{Modular flavor symmetries}
\label{sec:ModularFlavorSymmetries}

\subsubsection{Modular transformations of the wave functions $\wavefunction$}
\label{sec:ModularTransformationsOfWaveFunctions}

Crucially, physics should not depend on how we choose to parametrize the
underlying torus. That is, if we subject the half--period ratio $\tau$ of
the torus to a modular transformation, the physical predictions of the theory
have to stay the same. This means that there should be a dictionary between
theories with seemingly different but equivalent values of $\tau$, which are
related by modular transformations. 

Let us now study the action of $T$, under which $z \mapsto z$ and $\tau \mapsto
\tau + 1$. We wish to establish a dictionary between the wave functions on a
torus with parameter $\tau$ and an equivalent torus with parameter $\tau+1$. 
Let us now consider~\cite[equation~(37)]{Kikuchi:2020frp},
\begin{equation}\label{eq:Kikuchi:2020frp-37}
 \wavefunction(z,\tau,0)\xmapsto{~T~}
 \wavefunction(z,\tau+1,0)=
 \mathrm{e}^{\I\pi\frac{j^2}{|M|}}\,
 \wavefunction(z,\tau,0)\;.
\end{equation}
As shown in~\cite{Kikuchi:2020frp}, this relation holds for even units of
magnetic flux $M$. However, for odd $M$ a relation of the form
\begin{equation}\label{eq:TtrafoEvenM}
 \wavefunction(z,\tau+1,0)=\sum\limits_{j'=0}^{M-1}
 [\rho(T)]_{jj'}\,\wavefunction[j'](z,\tau,0)
\end{equation}
\emph{cannot} be true because according to \cref{eq:wrongboundary} both sides
have different periodicities. That is, on the left--hand side of the equality
\eqref{eq:Kikuchi:2020frp-37} we see a function that is supposed to be
``periodic'' under $z\mapsto z+\tau'$ whereas on the right--hand side the
function is supposed to be ``periodic'' under $z\mapsto z+\tau$. According to
\cref{eq:wrongboundary}, for odd $M$ only one of these ``periodicities'' can 
hold.

At first sight, this statement may appear odd. One might think that the zero
modes $\wavefunction$ form a basis of eigenmodes of the Dirac operator with
eigenvalue $0$. So one may expect that the transformed wave functions can be
expanded in terms of the original ones as in \cref{eq:TtrafoEvenM}. However,
this argument is incorrect. When we write down our wave functions we make a
choice for the origin of the torus. \emph{A priori} there are arbitrarily many choices
possible, which may be parametrized by $\Delta z$ in $\wavefunction(z + \Delta
z,\tau,0)$. So, on general grounds we only know that 
\begin{equation}\label{eq:TtrafoComplete}
 \wavefunction(z,\tau+1,0)=\sum\limits_{j'=0}^{M-1}
 [\rho(T)]_{jj'}\,\wavefunction[j'](z + \Delta z,\tau,0)
\end{equation}
for some appropriate real constant $\Delta z$. As we shall see, an appropriate
choice of $\Delta z$ will allow us to express the transformed wave functions in
terms of the original one also for odd $M$. More concretely, we will impose that
$z \mapsto z + \Delta z$, with some real constant $\Delta z$ that we are going
to find. Inserting this ansatz leads to (cf.\ \cref{eq:bshiftdetailed})
\begin{equation}
\wavefunction(z + \Delta z,\tau+1,0)= \widetilde{\mathcal{N}} \mathrm{e}^{\I\,\pi M \Delta z \frac{\im z}{\im\tau}} \mathrm{e}^{\I\,\pi M z \frac{\im z}{\im\tau}}
	\vartheta\orb{\frac{j}{M}}{0}(M(z+\Delta z+\nicefrac{1}{2} ) ,M\tau)\label{eq:bparm}
\end{equation}
Thus, if $N := M\left(\Delta z + \nicefrac{1}{2}\right)$ is an integer, we might
use \cref{eq:z+n}, which we recast here in a slightly different form 
\begin{align}
 \vartheta\orb{\frac{j}{M}}{0}(Mz+N,\tau)
 & =  
 \mathrm{e}^{2\pi\I\,N\,\alpha}\,
 \vartheta\orb{\frac{j}{M}}{0}(Mz,M\tau)
 \;,\label{eq:z+n2}
\end{align}
to get rid of the extra factor in the $z$ coordinate of the $\vartheta$
function. Finally, after the redefinition $z \mapsto z - \Delta z$, we obtain
\begin{align}\label{eq:oddMshift}
	\wavefunction(z,\tau,0)\xmapsto{~T~}
	\mathrm{e}^{\I\pi M \Delta z\frac{\im(z)}{\im\tau}}
	\,\mathrm{e}^{\I\pi\frac{j^2}{|M|}+2\I\pi j \Delta z}\,
	\wavefunction(z -\Delta z,\tau,0)\;.
\end{align}
Note that in order to get an integer $N$, it is sufficient to demand an integer
or half--integer $\Delta z$ for even $M$. For $\Delta z=0$ \cref{eq:oddMshift}
reproduces \cref{eq:Kikuchi:2020frp-37}. However, for odd $M$ we need a
half--integer $\Delta z$. Specifically, for $\Delta z = \nicefrac{1}{2}$ we find
that (see \cref{app:boundary_conditions_explicit_check} for details)
\begin{subequations}\label{eq:WaveFunctionTrafoSummary}
\begin{align}
 \wavefunction\left(z,\tau,0\right)
 \xmapsto{~S~}{}&\frac{\mathrm{e}^{\I \frac{\pi}{4}}}{\sqrt{M}}
 \bigg(-\frac{\tau}{|\tau |}\bigg)^{\nicefrac12}
 \sum_{k=0}^{M-1}\mathrm{e}^{2\pi \I jk/M}\,
 \psi^{k,M}\left(z,\tau,0\right)  \notag\\
 ={}&-\bigg(-\frac{\tau}{|\tau |}\bigg)^{\nicefrac12}\left[\rho({S})_{M}^{\psi}\right] _{jk}\psi^{k,M}(z,\tau,0)
 \;,\label{eq:WaveFunctionTrafoSummaryS}\\
 \wavefunction\left(z,\tau,0\right)
 \xmapsto{~T~}{}&
 \mathrm{e}^{\I\pi M\frac{\im z}{2\im\tau}}\, 
 \mathrm{e}^{\I\pi j(j/M + 1)} \,\wavefunction(z - \nicefrac12,\tau,0) \notag  \\
 ={}& \mathrm{e}^{\I\pi M\frac{\im z}{2\im\tau}}\left[\rho({T})_{M}^{\psi}\right]_{jk}\psi^{k,M}(z  -\nicefrac12,\tau,0)
 \;,\label{eq:WaveFunctionTrafoSummaryT} 
\end{align}
\end{subequations}
where 
\begin{subequations}\label{eq:WaveFunctionTrafoSummaryMatrices}
\begin{align}
\left[\rho(S)_{M}^{\psi}\right]_{jk}&:=-\frac{\mathrm{e}^{\I\pi/4}}{\sqrt{M}}\exp\left(\frac{2\pi\I\,j\,k}{M}\right)
 \;,\label{eq:WaveFunctionTrafoSummarySMatrix}\\
\left[\rho(T)_{M}^{\psi}\right]_{jk}&:=\exp\left[\I \pi\,j\left(\frac{j}{M}+1\right)\right]\delta_{jk}
 \;.\label{eq:WaveFunctionTrafoSummaryTMatrix} 
\end{align}
\end{subequations}
As we shall confirm shortly in \cref{eq:4Dfieldtransformations}, the
matrices~\eqref{eq:WaveFunctionTrafoSummaryMatrices} equal, up to a phase in
\cref{eq:WaveFunctionTrafoSummarySMatrix}, representation matrices of the
generators of finite metaplectic modular groups. They are compatible with
\cite{Kikuchi:2020frp,Kikuchi:2020nxn,Kikuchi:2021ogn,Hoshiya:2020hki}, but
\eqref{eq:WaveFunctionTrafoSummarySMatrix} differs from \cite{Ohki:2020bpo} by
the $\mathrm{e}^{\I \pi/4}$ phase. 
For even $M$, the $T$ transformation can rather be represented as in
\cref{eq:WaveFunctionTrafoSummary,eq:WaveFunctionTrafoSummaryMatrices} or
\cref{eq:Kikuchi:2020frp-37} due to the freedom of choosing half--integer or
integer $\Delta z$. However, since the Yukawa integral involves wave functions
with both odd and even fluxes $M$, we need to be consistent in our choice of 
$\Delta z$ to cancel the $z$--dependent phase appearing in
\cref{eq:WaveFunctionTrafoSummaryT} (cf.\ \cref{eq:YijkTrafoTDetailed3}).
Specifically, we need $\Delta z = \nicefrac12$ for the $T$ transformation also
for even $M$, in which case our results differ from
\cite{Ohki:2020bpo,Kikuchi:2020frp,Kikuchi:2020nxn,Hoshiya:2020hki} by phase
factors which are absent in \cref{eq:Kikuchi:2020frp-37}. Nevertheless, the
modular $T$ transformation of the 2D compact wave functions for odd $M$ was
excluded in \cite{Ohki:2020bpo,Kikuchi:2020frp,Kikuchi:2020nxn,Hoshiya:2020hki}.
In \cite{Kikuchi:2021ogn} they were introduced through the so--called
Scherk--Schwarz phases. In particular, our
\cref{eq:WaveFunctionTrafoSummary,eq:WaveFunctionTrafoSummaryMatrices} are
consistent with their discussion in \cite[equation~(126)]{Kikuchi:2021ogn}. 
However, as discussed in \cref{sec:mfs-even-odd}, we disagree with the statement
made in
\cite{Ohki:2020bpo,Kikuchi:2020frp,Kikuchi:2020nxn,Kikuchi:2021ogn,Hoshiya:2020hki}
that the modular transformed wave functions do not follow the appropriate
boundary conditions. As we have shown,  the $T$ transformation can generally
\emph{not} be represented by a matrix multiplication of the set of wave
functions, but necessarily goes beyond this. However, as we discuss in
\cref{app:T_transform_yukawa}, the extra exponential factors in
\cref{eq:WaveFunctionTrafoSummaryT} get canceled in the overlap integral
\eqref{eq:Yijk}, thus allowing us to define a matrix representation for the
transformation of the 4D fields, which derive from
\cref{eq:WaveFunctionTrafoSummaryMatrices}.

\subsubsection{Modular flavor symmetries in the effective 4D theory}
\label{sec:4dtheorysymmetries}

Let us now define proper ``modular flavor transformation'' for the 4D fields.
The first thing to notice is that these transformations \emph{cannot be unique},
at least not in models of this type.\footnote{It has been suggested that the
modular flavor symmetries can be defined by the requirement that the 6D fields
remain invariant \cite{Ohki:2020bpo}. However, apart from the fact that this
prescription fails for odd $M$ since the 2D coordinates gets shifted (cf.\ 
\cref{eq:WaveFunctionTrafoSummaryT}), it is not clear to us why one should
impose this very requirement.\label{ftn:6Dinverse4D}} The reason is that there
are additional symmetries at play, such as the remnant gauge factors, and we can
always add an extra transformation to our transformation law. That is to say
that the details of the representation matrices of a modular flavor
symmetries acting on the fields are somewhat ambiguous. Let us start with
something unambiguous: the transformation of the Yukawa couplings. As we have
seen in \cref{eq:explicit_Yukawa}, there are \emph{a priori} $\lambda$ Yukawa
couplings, out of which at most $\lambda/2+1$ are independent, as shown in
\cref{app:IndependentYukawas}.

Let us make an important distinction between ``physical Yukawa coupling''
$Y_{ijk}$ and ``holomorphic Yukawa couplings'' $\mathcal{Y}_{ijk}$
\cite{Kaplunovsky:1993rd}, which are related by (cf.\
\cite[equation~(5.41)]{Cremades:2004wa})
\begin{equation}\label{eq:RelationPhysicalVsHolomorphicYukawa}
 Y_{ijk}(\tau)=\mathrm{e}^{\widehat{K}/2}\,
 \frac{\mathcal{Y}_{ijk}(\tau)}{(K_{i\bar\imath}\,K_{j\bar\jmath}\,K_{k\bar k})^{\nicefrac12}}
 \;.
\end{equation}
Here, $\widehat{K}$ stands for the K\"ahler potential of the moduli, which is,
in our truncated setup, at tree level given by \cref{eq:KModuli}. The formula
for the Yukawa couplings~\eqref{eq:explicit_Yukawa}, which we obtained from the
overlap integral~\eqref{eq:Yijk}, contains the normalization factor
\eqref{eq:YukawaNormalization}, which is not holomorphic. In our case, the
matter field K\"ahler metric is proportional to $(\im\tau)^{-\nicefrac{1}{2}}$
(cf.\ \cref{eq:MatterFieldKaehlerMetric}), so  (cf.\
\cite[section~5.3]{Cremades:2004wa})
\begin{equation}
 \frac{\mathrm{e}^{\widehat{K}/2}}{(K_{i\bar\imath}\,K_{j\bar\jmath}\,K_{k\bar k})^{\nicefrac12}} 
 =\mathcal{N}_{abc}
 \propto g\,\left(\frac{\im\tau}{\mathcal{A}^2}\right)^{\nicefrac{1}{4}}\;.
\end{equation}
While $Y_{ijk}(\tau)$ is normalized and thus ``physical'', it is not holomorphic.
On the other hand, the superpotential coupling
\begin{equation}\label{eq:HolomorphicYukawaCoupling}
 \mathcal{Y}_{ijk}(\tau)=
 \vartheta\Orb{\widehat{\alpha}_{ijk}/\lambda}{0}(0,\lambda\,\tau)
\end{equation}
\emph{is} a proper modular form. Here, we have made use of the fact that the
upper characteristic is of the form $\widehat{\alpha}_{ijk}/\lambda$ with some
integer $\widehat{\alpha}_{ijk}$, cf.\ the discussion below
\cref{eq:explicit_Yukawa}, and we set, as done throughout this section, the
Wilson lines to zero. Further, all additional non--zero factors appearing in
\cref{eq:explicit_Yukawa} must be included in the K\"ahler potential, so that
they are canceled in the holomorphic couplings through the
redefinition~\eqref{eq:RelationPhysicalVsHolomorphicYukawa}.
The holomorphic coupling $\mathcal{Y}_{ijk}(\tau)$ differs from the
physical coupling between  canonically normalized fields by a non--holomorphic
factor. The modular transformations are seemingly non--unitary because of the
automorphy factor has generally not modulus $1$. However, the automorphy factors
get canceled, cf.\ our discussion below \cref{eq:ModularTransformation4DFields_1}. 

As shown in \cref{app:T_transform_yukawa}, the $\lambda$--plet of Yukawa
couplings transforms with the simple transformation law
\begin{equation}\label{eq:ModularTransformationYukawas_1}
 \mathcal{Y}_{\widehat{\alpha}}(\tau)\xmapsto{~\widetilde\gamma~}\mathcal{Y}_{\widehat{\alpha}}(\widetilde\gamma\,\tau)
 =\pm(c\,\tau+d)^{\nicefrac{1}{2}}\,\rho_{\rep{\lambda}}(\widetilde\gamma)_{\widehat{\alpha}\widehat{\beta}}\,\mathcal{Y}_{\widehat{\beta}}(\tau)\;,
\end{equation}
where $\widehat{\alpha}$ and $\widehat{\beta}$ are integers that label the
distinct Yukawa couplings, and we use the metaplectic element
$\widetilde{\gamma}\in\widetilde{\Gamma}$ instead of $\gamma\in\Gamma$ 
because the Yukawa couplings have weight $k_Y=\nicefrac{1}{2}$.
The transformation matrices of the modular generators are given by
\begin{subequations}\label{eq:Yukawas_Trafos_L}
\begin{align}
 \rho_{\rep{\lambda}}(\widetilde{S})_{\widehat{\alpha}\widehat{\beta}}&=-\frac{\mathrm{e}^{\I\pi/4}}{\sqrt{\lambda}}
 \,\exp\left(\frac{2\pi\I\,\widehat{\alpha}\,\widehat{\beta}}{\lambda}\right)\;,
 \label{eq:Yukawas_Trafos_L_S}\\
 \rho_{\rep{\lambda}}(\widetilde{T})_{\widehat{\alpha}\widehat{\beta}}&=\exp\left(\frac{\I\pi\,\widehat{\alpha}^2}{\lambda}\right)\,
 \delta_{\widehat{\alpha}\widehat{\beta}}\;.\label{eq:Yukawas_Trafos_L_T}
\end{align}
\end{subequations}
These matrices are symmetric and unitary, so that 
\begin{equation}\label{eq:ConjugateOfRepresentationMatrixIsInverse}
 [\rho_{\rep{\lambda}}(\widetilde S)_{\widehat{\alpha}\widehat{\beta}}]^{-1}=[\rho_{\rep{\lambda}}(\widetilde S)_{\widehat{\alpha}\widehat{\beta}}]^*
 \quad\text{and}\quad
 [\rho_{\rep{\lambda}}(\widetilde T)_{\widehat{\alpha}\widehat{\beta}}]^{-1}=[\rho_{\rep{\lambda}}(\widetilde T)_{\widehat{\alpha}\widehat{\beta}}]^*\;.
\end{equation}
Since there can be relations between the Yukawa couplings, this may not be an
irreducible representation. The relations between the Yukawa couplings depend on
the choice of fluxes. We will specify the irreducible representations of the
Yukawa couplings in our survey of models in \cref{sec:Models}. The modular
transformations of the Yukawa couplings given by \cref{eq:Yijk} were also
studied in \cite{Ohki:2020bpo}. Although an explicit general formula for any
combination of  $\mathcal{I}_{\alpha\beta}$ was not given in their work, our
results from \cref{eq:Yukawas_Trafos_L} match their result up to the phase
$\mathrm{e}^{\I \frac{\pi}{4}}$ in the models described in
\cref{sec:ToyModel_336,sec:ToyModel_224,sec:ToyModel_123}. This phase is crucial
to have the transformation matrices~\eqref{eq:Yukawas_Trafos_L} satisfy the
presentation~\eqref{eq:tildeGamma_4N-presentation} and, thus, give rise to
representations of a finite metaplectic modular group, as was noted in
\cite{Hoshiya:2020hki}.  Note also that there is an extra minus in our
\cref{eq:Yukawas_Trafos_L} compared to \cite[equations~(64) and
(108)]{Ohki:2020bpo} and \cite{Hoshiya:2020hki}. However, this sign comes only
from our convention that the automorphy factor is
$\varphi(S,\tau)=-\sqrt{-\tau}$ in \cref{eq:MetaplecticTrafof}.

Next, we discuss modular flavor symmetries. They are, by definition, symmetry
transformations of the 4D Lagrange density. In our present discussion, we are
thus seeking transformations of the 4D fields, $\phi^{j,M}$, which are such that
superpotential couplings
\begin{equation}
 \mathscr{W}\supset \mathcal{Y}_{ijk}(\tau)\,\phi^{i,\mathcal{I}_{ab}}
 \,\phi^{j,\mathcal{I}_{ca}}\,\phi^{k,\mathcal{I}_{cb}}
\end{equation}
are invariant up to K\"ahler transformations, cf.\ the discussion around
\eqref{eq:AutomorphyW}. Here,
$\mathcal{I}_{cb}=-\mathcal{I}_{bc}=\mathcal{I}_{ab}+\mathcal{I}_{ca}>0$. That
is, our modular flavor transformations are given by
\begin{equation}\label{eq:ModularTransformation4DFields_1}
  \phi^{j,M}\xmapsto{\widetilde\gamma~}\pm(c\,\tau+d)^{-\nicefrac12}\,\left[\rho_{M}^\phi(\widetilde\gamma)\right]_{jk}^{-1}\,\phi^{k,M}\;.
\end{equation}
Notice that, due
to~\cref{eq:ModularTransformationYukawas_1,eq:ModularTransformation4DFields_1}, 
the superpotential acquires modular weight $k_\mathscr{W}=-1$, see
\cref{eq:AutomorphyW}. The corresponding automorphy factor gets canceled by the
transformation of $\tau$ in the K\"ahler potential followed by a K\"ahler
transformation, see our discussion around~\cref{eq:ModularTrafoTau-TauBar}.
Therefore, the requirement that the modular transformations be a symmetry
amounts to demanding that
\begin{multline}
 \mathcal Y_{ijk}(\widetilde\gamma\,\tau)\,
 \left[\rho_{\rep{\mathcal{I}_{ab}}}^\phi(\widetilde\gamma)\right]_{ii'}^{-1}\phi^{i',\mathcal{I}_{ab}}\,
 \left[\rho_{\rep{\mathcal{I}_{ca}}}^\phi(\widetilde\gamma)\right]_{jj'}^{-1}\phi^{j',\mathcal{I}_{ca}}\,
 \left[\rho_{\crep{\mathcal{I}_{cb}}}^\phi(\widetilde\gamma)\right]_{kk'}^{-1}\phi^{k',\mathcal{I}_{cb}}
 \\
 \stackrel{!}{=}
 \mathcal Y_{ijk}(\tau)\,\phi^{i,\mathcal{I}_{ab}}
 \,\phi^{j,\mathcal{I}_{ca}}\,\phi^{k,\mathcal{I}_{cb}}\;.
\end{multline}
As already mentioned, this condition does not fix the transformation laws of the
4D fields uniquely. However, we can use the transformation properties of the
$\mathds{T}^2$ wave functions, (cf.\
\cref{eq:YijkTrafoTDetailed3,eq:YijkTrafoSDetailed}), to infer the matrix
structure of the transformations. One way in which we may infer the
transformations of the 4D fields is by using the \emph{quasi}--inverse
transformations of the compact wave functions, that is, the inverse
transformations of \cref{eq:WaveFunctionTrafoSummaryMatrices}. However, a more
convenient choice is   
\begin{subequations}\label{eq:4Dfieldtransformations}
\begin{align}
 \rho_{\rep{M}}^\phi(\widetilde S)_{jk}&=-\frac{\mathrm{e}^{\I\pi(3M+1)/4}}{\sqrt{M}}
 \,\exp\left(\frac{2\pi\I\,j\,k}{M}\right)\;,\\
 \rho_{\rep{M}}^\phi(\widetilde T)_{jk}&=\exp\left[\I\pi\,j\left(\frac{j}{M}+1\right)\right]\,
 \delta_{jk}\;,
\end{align}
\end{subequations}
where we have chosen the
transformation~\eqref{eq:WaveFunctionTrafoSummarySMatrix}  multiplied by a phase
$\mathrm{e}^{3\I\pi\frac{M}{4}}$ in the $S$ matrix representation.  These
matrices fulfill \cref{eq:ConjugateOfRepresentationMatrixIsInverse}, too. This
choice has the virtue that
$\rho_{\crep{M}}(\widetilde{\gamma})=[\rho_{\rep{M}}(\widetilde{\gamma})]^*$
and that, as we will demonstrate in \cref{sec:Models}, it yields the correct
representation matrices for the group $\widetilde{\Gamma}_{2\lambda}$.

We also note that, as far as the Yukawa couplings are concerned, there is a
$\U1$ symmetry due to the condition of \cref{eq:FluxCondition}, which acts
as
\begin{equation}
\phi^{j,\mathcal{I}_{\alpha\beta}}\xmapsto{~\U1~}
\mathrm{e}^{\I q \alpha\,\mathcal{I}_{\alpha\beta}}
\phi^{j,\mathcal{I}_{\alpha\beta}}\;,
\label{eq:fluxcondition4Dfields}
\end{equation}
where $\alpha,\beta\in\{a,b,c\}$ as in \cref{eq:I_alpha_beta}.
Here, $\phi^{\mathcal{I}_{ab}},\phi^{\mathcal{I}_{ca}}$ have a charge $+1$ and
$\phi^{\mathcal{I}_{cb}}$ a charge $-1$. This \U1 factor allows one to install
``extra'' phases of the above type. Note that while the $T$--transformed wave
functions, for odd $M$, cannot be expanded in terms of untransformed wave
functions, the additional factor in our dictionary
\eqref{eq:WaveFunctionTrafoSummaryT} cancels in the overlap integrals
\eqref{eq:Yijk} so that there is a meaningful, well--defined modular flavor
transformation of the 4D fields also for odd $M$.  Our proposal in
\cref{eq:4Dfieldtransformations} for the transformations of 4D fields
$\phi^{j,M}$ for even values of $M$ differs from the results in
\cite{Ohki:2020bpo}. While \cite{Ohki:2020bpo} assumes that the modular
transformations of the 4D fields coincide with those of the 2D wave functions,
we assume the 4D fields transform quasi--inversely to the 2D wave functions.
Furthermore, we have an extra phase $\mathrm{e}^{3\I\pi \frac{M}{4}}$, which
is useful to achieve metaplectic group representations.
Note that we specify the $T$ transformation, rather than just the $T^2$
representation as in \cite{Ohki:2020bpo}.

\subsection{Models}
\label{sec:Models}

In this subsection, we survey a couple of toy models. These models are far from
realistic but highlight how modular flavor symmetries derive from some simple
magnetized tori with even and odd numbers of repetitions of matter fields. In
all of the next models we will use the representation matrices stated in
\cref{eq:4Dfieldtransformations} for the $\mathcal{I}_{cb}$--plet of $\phi^{k}$
4D fields, while the $\mathcal{I}_{ab}$--plet of $\phi^{i}$ and
$\mathcal{I}_{ca}$--plet of $\phi^{j}$ 4D fields will transform in the conjugate
representation. On the other hand, the $\lambda$--plet of Yukawa couplings
will follow the representation matrices found in~\cref{eq:Yukawas_Trafos_L}. We
will show that the modular flavor symmetries in these models are given by
$\widetilde{\Gamma}_{2\lambda}$ with $\lambda$ being the least common multiple
of matter repetition numbers~\eqref{eq:def_L_lcm}. Furthermore, using
\cref{eq:P=lambdad2} one can see that for a fixed total number of Yukawa
couplings $P$, the largest number of independent Yukawa couplings, that is the
largest $\lambda$, is obtained by having the least possible $d$. Although we
have proposed the representation matrices for the 4D fields in
\cref{eq:4Dfieldtransformations}, the ones for the Yukawa couplings
\cref{eq:Yukawas_Trafos_L} are unambiguous. In fact, in all models we discuss
here we will find that the representations  $\rho_{\rep\lambda}$
satisfy~\cref{eq:tildeGamma_4N-presentation} together with the finiteness
conditions~\eqref{eq:tildeGamma_4(2)-presentation}--\eqref{eq:tildeGamma_4(3)-presentation} 
for $N=2,3$, with $\lambda=2N$. Thus, the modular transformations of the
Yukawa couplings build representations of the finite metaplectic group
$\widetilde\Gamma_{2\lambda}$. In \cite{Hoshiya:2020hki} it was also noted that,
for even numbers of flavors, the Yukawa couplings transform as a $\lambda$--plet
under the metaplectic group. However, in \cite{Hoshiya:2020hki} it does not get
mentioned that for $\lambda>2$ this representation is reducible, which is rather
easy to see from our general compact expression \eqref{eq:explicit_Yukawa}, but
less obvious when one represents the Yukawa coupling as the sum
\eqref{eq:CIM5.12}. Moreover, we will demonstrate in each model that,
independently of whether the number of flavors is even or odd, the
transformations of the 4D fields encoded in
$\rho_{\rep{\mathcal{I}_{\alpha\beta}}}^\phi$ build representations of the same
group, so that $\widetilde{\Gamma}_{2\lambda}$ can be regarded as the modular
flavor symmetry of the models. We are hence led to conjecture that, with
$\lambda$ from \cref{eq:def_L_lcm},
\begin{equation}\label{eq:conjecture}
\text{magnetized tori with $\lambda=\lcm(\text{\# of flavors})$ exhibit a 
  $\widetilde{\Gamma}_{2\lambda}$ modular flavor symmetry}\;.
\end{equation}

\subsubsection{Model with $\mathcal{I}_{ab}=\mathcal{I}_{ca}=1$ and $\mathcal{I}_{bc}=-2$}
\label{sec:ToyModel_112}

Let us consider a model based on a $\U{3}$ gauge symmetry and fluxes
\begin{align}
	F &= \frac{\pi \I}{\im\tau}\begin{pmatrix} 
	0 & \hphantom{-}0 & 0 \\
    0 &-1 & 0 \\
    0 & \hphantom{-}0 & 1 \\
    \end{pmatrix}\;.
\end{align}
The fluxes break $\U{3}\rightarrow\U1_a\times\U1_b\times\U1_c$. Since the
$N_\alpha=1$ for $\alpha\in\{a,b,c\}$, we thus have  
\begin{equation}
 \mathcal{I}_{ab}=\mathcal{I}_{ca}=1\quad\text{and}\quad\mathcal{I}_{bc}=-2\;.
\end{equation}
According to \eqref{eq:I_alpha_beta} this means that we have one repetition of
$\wavefunction[i,\mathcal{I}_{ab}=1]$ and $\wavefunction[j,\mathcal{I}_{ca}=1]$
each, and two copies of $\wavefunction[k,\mathcal{I}_{bc}=-2]$. We can now
compute the holomorphic Yukawa couplings of this model using \eqref{eq:explicit_Yukawa},
\begin{align}
 \mathcal Y_{ijk}(0,\tau) &= \vartheta\orb{\frac{k}{2}}{0}\left(0,2\,\tau\right)\;,
\end{align}
which gives a doublet\footnote{Here, we use the notation $\rep{\widehat{2}}$
from \cite{Liu:2020msy} to refer to the two--dimensional irreducible
representation of $\widetilde{\Gamma}_4$.}
\begin{align}
\mathcal Y_{\rep{\widehat{2}}}=\begin{pmatrix}
\mathcal Y_{0}\\
\mathcal Y_{1}\\
\end{pmatrix}
:= \begin{pmatrix}
\mathcal Y_{000}\\
\mathcal Y_{001}\\
\end{pmatrix}\; ,
\end{align}
which transforms under the representation matrices given by
\begin{equation}\label{eq:rho_Y_2}
\rho_{\rep{\widehat{2}}}(\widetilde S)= \frac{-\mathrm{e}^{\frac{\I \pi }{4}}}{\sqrt{2}}
\begin{pmatrix}
 1 & \hphantom{-}1 \\
1 & -1 \\
\end{pmatrix}
\quad\text{and}\quad
\rho_{\rep{\widehat{2}}}(\widetilde T)=
\begin{pmatrix}
 1 & \hphantom{-}0 \\
 0 & \hphantom{-}\I \\
\end{pmatrix}\;.
\end{equation}
Note that in this case we could have used
\cite[equation~(5.17)]{Cremades:2004wa} since $\mathcal{I}_{ab}=d=1$. This
Yukawa coupling coincides (up to an irrelevant similarity transformation with
$\diag(1,-1)$) with the $\boldsymbol{\widehat{2}}$ representation of
$\widetilde{\Gamma}_{2 \lambda=4}=\widetilde{S}_4\cong[96,67]$ \cite[cf.\
equation~(41)]{Liu:2020msy}. This representation can be thought of as the
fundamental representation of $\widetilde{S}_4$ in that all other nontrivial
representations can be obtained by reducing tensor products of a suitable number
of $\boldsymbol{\widehat{2}}$ representations. The fields with multiplicity 2
transform with the inverses (or conjugates, see
\cref{eq:ConjugateOfRepresentationMatrixIsInverse}) of the representation
matrices~\eqref{eq:rho_Y_2}. That is, these fields transform under 
$\rep{\overline{\widehat{2}}}$. Altogether, we have a $\widetilde{S}_4$ theory
with (holomorphic) Yukawa couplings given by
\begin{equation}
 \mathscr{W}\supset 
 \phi_{ab}\,\phi_{ca}\,\left(
  \vartheta\orb{0}{0}\left(0,2\,\tau\right)\,\phi_{cb}^0
  +\vartheta\orb{\nicefrac{1}{2}}{0}\left(0,2\,\tau\right)\,\phi_{cb}^1
 \right)\;,
\end{equation}
where we suppress the trivial generation indices of the fields coming with
repetition $1$. Notice that the physical Yukawa coupling comes with extra
normalization factors, see \cref{eq:RelationPhysicalVsHolomorphicYukawa}.

\subsubsection{Model with $\mathcal{I}_{ab}=\mathcal{I}_{ca}=3$ and $\mathcal{I}_{bc}=-6$}
\label{sec:ToyModel_336}

Let us consider a three generation toy model, based on a super--Yang--Mills
theory in six dimensions  with gauge group \U4~\cite{Ohki:2020bpo}. The two extra dimensions are
compactified on $\mathds{T}^2$, and the \U4 gauge symmetry gets broken to
$\SU2\times\U1_a\times\U1_b\times\U1_c$ by the fluxes
\begin{align}
	F &= \frac{\pi \I}{\im\tau}
	\begin{pmatrix} 
	\mathds{1}_{2\times 2} & 0 & 0 \\
    0 & -3 & 0 \\
    0 & 0 & 3 \\
    \end{pmatrix}\;,
\end{align}
where we used \cref{eq:fluxF}. The chiral matter content of the supersymmetric
model is given in \cref{tab:336ModelMatter}. They decompose into three
generations of $L$ particles, six generations of $R$ particles and three
generations of $H$ particles.
\begin{table}[t!]
\centering
\begin{tabular}{lcl}
\toprule
field       & \begin{tabular}[b]{@{}c@{}}$\SU2\times\U1_a\times\U1_b\times\U1_c$\\
 quantum numbers\end{tabular}  & \# of copies                  \\ 
\midrule
$L$  & $\rep{2}_{(1,-1,0)}$
& $\mathcal{I}_{ab}=2-(-1)=3$   \\
$R$ & $\rep{1}_{(0,+1,-1)}$
& $\mathcal{I}_{bc}=-1-(5)=-6$  \\
$H$   & $\rep{2}_{(-1,0,1)}$
& $\mathcal{I}_{ca}=5-(2)=3$ \\
\bottomrule
\end{tabular}
\caption{Matter content of the 336 model.}
\label{tab:336ModelMatter}
\end{table}
The superpotential of this model is given by 
\begin{equation}\label{eq:Superpotential336Model}
\mathscr{W}\supset \mathcal Y_{ijk}L^iH^jR^k\;,
\end{equation}
where the (holomorphic) Yukawa couplings are given by \cref{eq:explicit_Yukawa},
\begin{align}
    \mathcal Y_{ijk}(\tau ) &= 
	\vartheta\orb{\frac{k-2j}{6}}{0}(0,6\tau)
	\;.\label{eq:Yijk_336}
\end{align}
Here we used the values from \cref{tab:336ModelMatter} and assumed zero
Wilson lines. The explicit transformation matrices for the $L^i$ and $H^j$ are
given by \eqref{eq:4Dfieldtransformations} for $M=3$, and are the conjugates of
\begin{equation}\label{eq:rho_phi_3}
\rho_{\rep{3}}^\phi(\widetilde S)= -\frac{1}{\sqrt{3}}
\begin{pmatrix}
 \I & \I & \I \\
\I & \mathrm{e}^{-\frac{5 \I \pi }{6}} & 
\mathrm{e}^{-\frac{\I \pi }{6}} \\
\I & \mathrm{e}^{-\frac{\I \pi }{6}} & 
\mathrm{e}^{-\frac{5 \I \pi }{6}}\\
\end{pmatrix}
\quad\text{and}\quad
\rho_{\rep{3}}^\phi(\widetilde T)=\begin{pmatrix}
 1 & 0 & 0 \\
 0 & \mathrm{e}^{-\frac{2 \I \pi }{3}} & 0 \\
 0 & 0 & \mathrm{e}^{-\frac{2 \I \pi }{3}} \\
\end{pmatrix}\;.
\end{equation}
The explicit transformation matrices for the $R^{k}$ fields are given by  
\cref{eq:4Dfieldtransformations} for $M=6$
\begin{subequations}\label{eq:rho_phi_6}
\begin{align}
\rho_{\rep{6}}^\phi(\widetilde S)&= -\I\,\frac{\mathrm{e}^{\I \frac{\pi}{4}}}{\sqrt{6}}
\begin{pmatrix}
 1 & 1 & 1 & 1 & 1 & 1 \\
 1 & \mathrm{e}^{\frac{\pi\I }{3}} & \mathrm{e}^{\frac{2 \pi\I }{3}} & -1 & \mathrm{e}^{-\frac{2 \pi\I }{3}} & \mathrm{e}^{-\frac{\pi\I }{3}} \\
 1 & \mathrm{e}^{\frac{2 \pi\I }{3}} & \mathrm{e}^{-\frac{2 \pi\I }{3}} & 1 & \mathrm{e}^{\frac{2 \pi\I }{3}} & \mathrm{e}^{-\frac{2 \pi\I }{3}} \\
 1 & -1 & 1 & -1 & 1 & -1 \\
 1 & \mathrm{e}^{-\frac{2 \pi\I }{3}} & \mathrm{e}^{\frac{2 \pi\I }{3}} & 1 & \mathrm{e}^{-\frac{2 \pi\I }{3}} & \mathrm{e}^{\frac{2 \pi\I }{3}} \\
 1 & \mathrm{e}^{-\frac{\pi\I }{3}} & \mathrm{e}^{-\frac{2 \pi\I }{3}} & -1 & \mathrm{e}^{\frac{2 \pi\I }{3}} & \mathrm{e}^{\frac{\pi\I }{3}} \\
\end{pmatrix}\;,\\ 
\rho_{\rep{6}}^\phi(\widetilde T)&=
\diag\bigl(1,\mathrm{e}^{-\frac{5\pi\I }{6}},
\mathrm{e}^{\frac{2 \pi\I}{3}},\I,
\mathrm{e}^{\frac{2 \pi\I }{3}},\mathrm{e}^{-\frac{5\pi\I }{6}}\bigr)\;.
\end{align}
\end{subequations}
As discussed at the end of \cref{sec:YukawaCouplings}, there are only
$\nicefrac{\lambda}{2}+1=4$ independent Yukawa couplings,\footnote{The relations
given in \cref{eq:Yukawas_sixplet} are valid for both holomorphic and
non--holomorphic Yukawa couplings.}
\begin{subequations}\label{eq:Yukawas_sixplet}
\begin{align}
 Y_0&:= Y_{i=j,j,k=2j}\;, & Y_1&:= Y_{i=j+1,j,k=2j+1}= Y_5:= Y_{i=j+2,j,k=2j+5}\;, 
 \\
 Y_3&:= Y_{i=j,j,k=2j+3}\;,&
 Y_2&:= Y_{i=j+2,j,k=2j+2}= Y_4:= Y_{i=j+1,j,k=2j+4}\;,
\end{align}
\end{subequations}
where $i$ and $j$ are understood to be modulo $3$, and $k$ modulo $6$. The
six--plet of holomorphic Yukawa coupling coefficients  $\mathcal
Y_{\rep{6}}=(\mathcal Y_{0}, \mathcal Y_{1}, \mathcal Y_{2}, \mathcal Y_{3},
\mathcal Y_{4}, \mathcal Y_{5})^{T}$  obeys the transformation law 
\cref{eq:ModularTransformationYukawas_1} under modular transformations,
with the matrix representations
\begin{equation}\label{eq:rho_Y_6}
\rho_{\rep{6}}(\widetilde S)= -\I\, \rho_{\rep{6}}^\phi (\widetilde S)
\quad\text{and}\quad
\rho_{\rep{6}}(\widetilde T)=
\diag\bigl(1,\mathrm{e}^{\frac{\pi\I }{6}},
\mathrm{e}^{\frac{2 \pi\I}{3}},-\I,
\mathrm{e}^{\frac{2 \pi\I }{3}},\mathrm{e}^{\frac{\pi\I }{6}}\bigr)\;.
\end{equation}
However, the $6\times6$ matrices can be reduced to a 4--dimensional
representation due to the relation between the Yukawa couplings in
\cref{eq:Yukawas_sixplet}. Using the projection matrix
\begin{align}\label{eq:ProjectionMatrix6to4}
P_{6\to 4}&=
\begin{pmatrix}
 1 & 0 & 0 & 0 \\
 0 & \frac{1}{\sqrt{2}} & 0 & 0 \\
 0 & 0 & \frac{1}{\sqrt{2}} & 0 \\
 0 & 0 & 0 & 1 \\
 0 & 0 & \frac{1}{\sqrt{2}} & 0 \\
 0 & \frac{1}{\sqrt{2}} & 0 & 0 \\
\end{pmatrix}\; ,
\end{align}
we can define the $\rep{4}$--plet of independent Yukawa couplings through $\mathcal
Y_{\rep{4}}=P_{6\to 4}^{T}\mathcal Y_{\rep{6}}$, which transform as modular
forms with the representation matrices given by
\begin{subequations}\label{eq:rho_Y_6to4}
\begin{align}
\rho_{\rep{4}}(\widetilde S)&= P_{6\to 4}^{T}\rho_{\rep{6}}(\widetilde S)P_{6\to 4}
= -\,\frac{\mathrm{e}^{\I \frac{\pi}{4}}}{\sqrt{6}}
\begin{pmatrix}
 1 & \sqrt{2} & \sqrt{2} & 1 \\
 \sqrt{2} & 1 & -1 & -\sqrt{2} \\
 \sqrt{2} & -1 & -1 & \sqrt{2} \\
 1 & -\sqrt{2} & \sqrt{2} & -1 \\
\end{pmatrix}\;,\quad\text{and}\quad \\ 
\notag \\ 
\rho_{\rep{4}}(\widetilde T)&= P_{6\to 4}^{T}\,\rho_{\rep{6}}(\widetilde T)\,P_{6\to 4}
=\begin{pmatrix}
 1 & 0 & 0 & 0 \\
 0 & \mathrm{e}^{\frac{\pi\I }{6}} & 0 & 0 \\
 0 & 0 & \mathrm{e}^{\frac{2 \pi\I }{3}} & 0 \\
 0 & 0 & 0 & -\I \\
\end{pmatrix}\;.
\end{align}
\end{subequations}
The representation matrices in
\cref{eq:rho_phi_3,eq:rho_phi_6,eq:rho_Y_6,eq:rho_Y_6to4} fulfill the conditions
\eqref{eq:tildeGamma_4N-presentation} for $N=3$ and
\eqref{eq:tildeGamma_4(3)-presentation}, which implies that this model exhibits
a $\widetilde{\Gamma}_{2 \lambda=12}$ finite modular symmetry of order 2304. The
fact that there are only four distinct Yukawa entries implies that the
6--dimensional representation of the Yukawa  couplings decomposes into
$\widetilde{\Gamma}_{12}$ irreducible representations according  to
$\rep{6}=\rep{4}\oplus\rep{2}$, as we have confirmed, where  the doublet
vanishes.  As we shall discuss below, this can be also attributed to the
existence of an outer automorphism.  Using the character tables (cf.\
\cite[section~3.4]{Ramond:2010zz}), we find that the matter triplets and
six--plets are reducible as well, $\rep{3}=\rep{2}''\oplus\rep{1}'$ and 
$\rep{6}'=\rep{4}'\oplus\rep{2}'$, where we added primes to indicate that these
are different representation matrices, and that the singlet is nontrivial. We
have  verified that the reducible representation $\rep{6}'$ provides us with a
faithful  representation content of $\widetilde{\Gamma}_{12}$ and its tensor
products yield all other representations of the group. The six
$\mathcal{Y}_{\widehat{\alpha}}$ have been identified in \cite{Ohki:2020bpo},
where they have been represented as sums of three different
$\vartheta$--functions each, and the relations $Y_1=Y_5$ and $Y_2=Y_4$ have been
missed. The latter relations are actually quite interesting as they can be
thought of as $i\leftrightarrow j$ exchange symmetries,
\begin{equation}
 Y_{ijk}(\tau)=Y_{jik}(\tau)\;.
 \label{eq:exchangesymmetries}
\end{equation}
However, the wave functions labeled $i$ and $j$, i.e.\ the $L^i$ and $H^j$, have
\emph{different} quantum numbers in 4D (and in the upstairs theory). This means
that this symmetry is not an ``ordinary'' flavor symmetry but an outer
automorphism of the low--energy gauge symmetry. Note that the existence of
this very outer automorphism depends on the specifics of the model, i.e.\ while
both the current model and the one presented in \cref{sec:ToyModel_123} have a
$\widetilde{\Gamma}_{2\lambda}=\widetilde{\Gamma}_{12}$ metaplectic flavor
symmetry, the form of the outer automorphism is specific to the current model.
Examples for such outer automorphisms include the so--called left--right parity
\cite{Mohapatra:1979ia}. It is known that such symmetries can originate as
discrete remnants of gauge symmetries either by dialing appropriate
VEVs~\cite{Kibble:1982dd,Chang:1983fu} or by orbifolding
\cite{Biermann:2019amx}. As the exchange of the \U1 factors is part of the
original \U4 gauge symmetry of the model, we have identified yet another way in
which these outer automorphism can emerge from an explicit gauge symmetry.

\paragraph{A geometric interpretation of Yukawa couplings.}
It is instructive to discuss the geometrical interpretation of these results. We
have derived the couplings by computing the overlaps of wave functions, see
\eqref{eq:Yijk}. The result is that, up to a normalization factor, the Yukawa
couplings are given by
\begin{equation}\label{eq:YukawasAsOverlapsOfGaussians}
 Y_{\widehat{\alpha}}\propto(\im\tau)^{-1/4}\,\vartheta\Orb{\nicefrac{\widehat{\alpha}}{\lambda}}{0}(0,\lambda\,\tau)
 =(\im \tau)^{-1/4}\sum\limits_{\ell=-\infty}^\infty
 \mathrm{e}^{-\pi\,\lambda\left(\im\tau-\I\,\re\tau\right)(\nicefrac{\widehat{\alpha}}{\lambda}+\ell)^2}
 \;,
\end{equation}
where we have used \cref{eq:ThetaSum1}. Here we choose to highlight the fact that
the terms are exponentially suppressed by
$\mathrm{e}^{-\pi\,\lambda\,\im\tau\,\xi}$ with some $\xi>0$ in order to compare
our result for the Yukawa couplings with a simple overlap of Gaussians. For
simplicity, we just consider two Gaussians, and consider
\begin{equation}
 y(a,b_1,b_2)
 =\int\limits_{-\infty}^\infty\!\D x\,
 	\mathcal{N}_{b_1}\,\mathrm{e}^{-x^2/b_1}\,
	\mathcal{N}_{b_2}\,\mathrm{e}^{-(x-a)^2/b_2}
 =\frac{\mathrm{e}^{-a^2/(b_1+b_2)}}{\sqrt{\pi}\sqrt{b_1+b_2}}
\end{equation}
with Gaussian normalization factors $\mathcal{N}_b=1/\sqrt{b\pi}$. In order to
compute the overlap on the torus, one does not only have to compute the overlap
of a given Gaussian of width $b_1$, say, with one Gaussian of width $b_2$, but
with all images of the second Gaussian under torus translations. This leads to
an expression which is qualitatively similar to the sum on the right--hand side
of \cref{eq:YukawasAsOverlapsOfGaussians}.

\begin{figure}[t!]
\centering
 \includegraphics{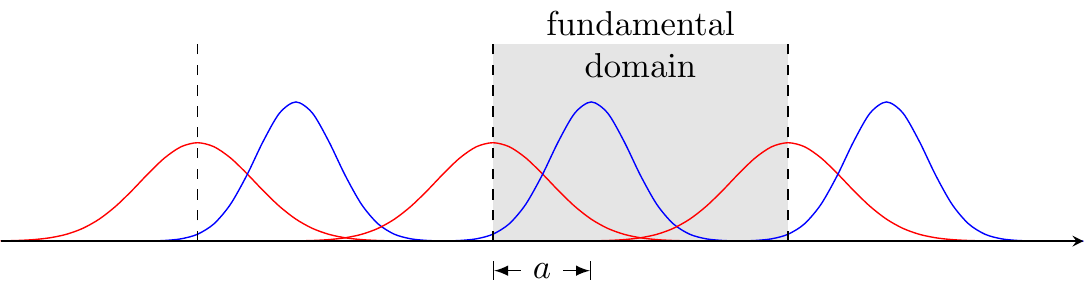}
 \caption{Overlap of two Gaussians on a torus. The overlap of a given, say
  red, curve is not just the overlap with one blue curve but with infinitely many
  of them, thus leading to an expression of the form 
  \eqref{eq:YukawasAsOverlapsOfGaussians}.}
 \label{fig:OverlapGaussian} 
\end{figure}

\begin{figure}[t!]
\centering
 \includegraphics[width=0.7\linewidth]{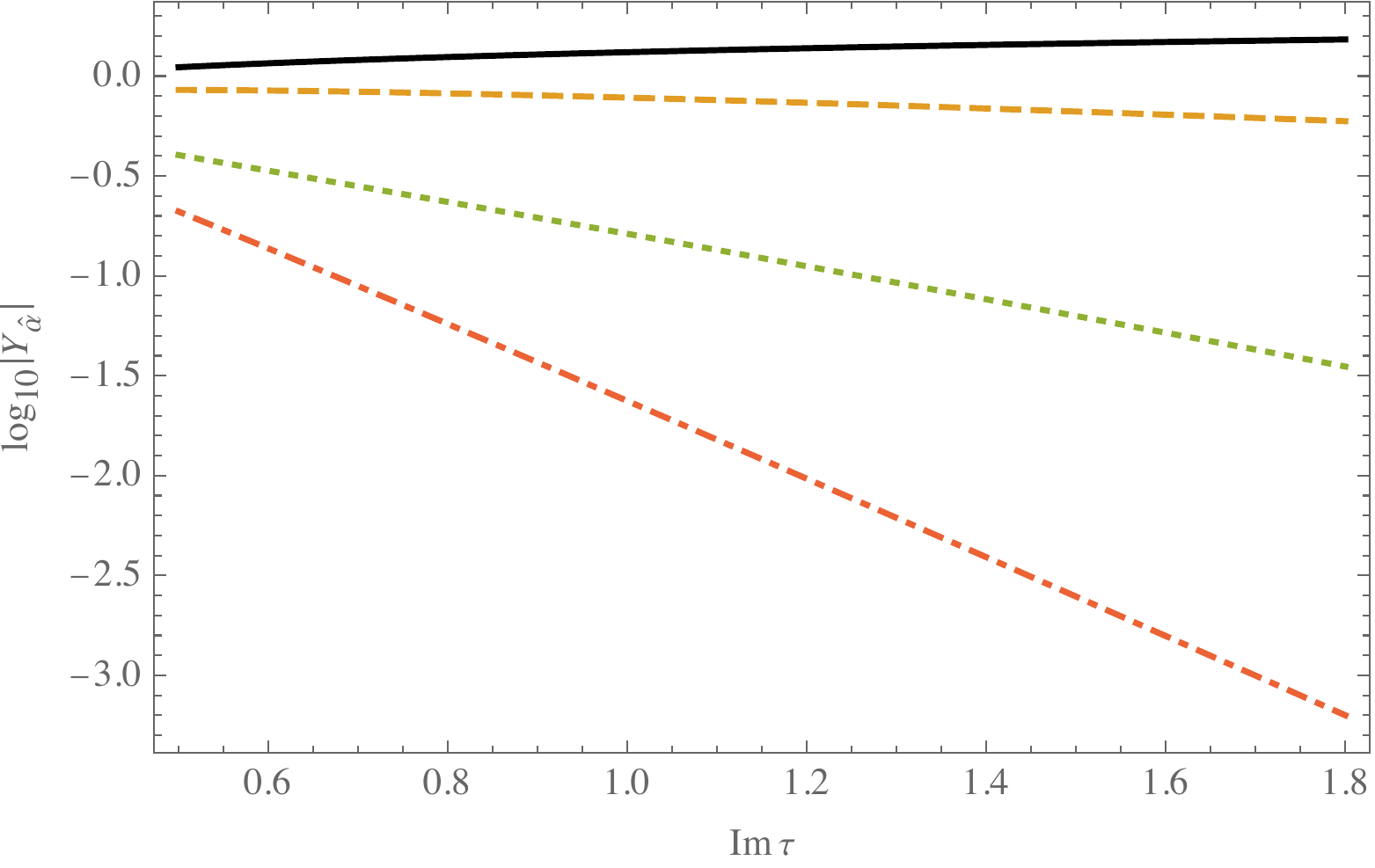}
 \caption{Dependence of the magnitude of the Yukawa couplings $Y_{\widehat{\alpha}}$ for
  $\re\tau=0.1$. The black solid, orange dashed, green dotted and red
  dash--dotted curves represent  $\widehat{\alpha}=0$, $1$, $2$ and $3$, respectively. 
  There is an exponential suppression with $\im\tau$ that depends on the
  ``distance'' between the wave functions $\widehat{\alpha}$, i.e.\ the $\im\tau$
  dependence is more pronounced for larger $\widehat{\alpha}$.}
 \label{fig:YukawaMagnitude} 
\end{figure}

Turning this around, the upper characteristics $\widehat{\alpha}$ in
\cref{eq:explicit_Yukawa,eq:YukawasAsOverlapsOfGaussians}, or, more precisely,
$\min\bigl(|\widehat{\alpha}/\lambda|,|1-\widehat{\alpha}/\lambda|\bigr)$ with
$0\le|\widehat{\alpha}/\lambda|<1$, has the interpretation of a ``distance between the
loci of the states'', i.e.\ $a$ in \cref{fig:OverlapGaussian}. We illustrate
this by plotting some sample Yukawa couplings in \cref{fig:YukawaMagnitude}.
This geometric intuition may conceivably provide us with an understanding of the
observed hierarchies of fermion masses. Apart from the fact that the kinetic
terms are under control, the geometric interpretation may be one of the
strongest motivations for deriving the modular flavor symmetries from explicit
tori. 

\subsubsection{Model with $\mathcal{I}_{ab}=\mathcal{I}_{ca}=2$ and $\mathcal{I}_{bc}=-4$}
\label{sec:ToyModel_224}

Although we have stressed that our results are valid for odd repetitions of
matter $\mathcal{I}_{\alpha\beta}$, they of course also apply to settings in
which all $\mathcal{I}_{\alpha\beta}$ are even. Let us consider a toy model with
the same superpotential and gauge group breaking as in \cref{sec:ToyModel_336}
by the fluxes
\begin{align}
	F &= \frac{\pi \I}{\im\tau}
	\begin{pmatrix} 
	\mathds{1}_{2\times 2} & 0 & 0 \\
    0 & \mathllap{-}2 & 0 \\
    0 & 0 & 2 \\
    \end{pmatrix}\;,
\end{align}
where we used \cref{eq:fluxF}. This means that we have two repetitions of
$L^{i}$ and $H^{j}$ each, and four copies of $R^{k}$. Similarly to what we have
found in \cref{sec:ToyModel_336}, there are $\lambda/2+1=3$ independent Yukawa
couplings,
\begin{subequations}\label{eq:Yukawas_fourplet}
\begin{align}
 Y_0&:=Y_{i=j,j,k=2j}\;, & Y_1&:=Y_{i=j+1,j,k=2j+1}=Y_3:=Y_{i=j+1,j,k=2j+3}\;,
 \\
 Y_2&:=Y_{i=j,j,k=2j+2} \;,
\end{align}
\end{subequations}
where $i$ and $j$ are understood to be modulo $2$, and $k$ modulo $4$. The
equality $Y_{1}=Y_{3}$ is also a consequence of the exchange symmetry given in
\cref{eq:exchangesymmetries}. The four--plet of holomorphic Yukawa
couplings $\mathcal Y_{\widehat{\alpha}}$ follow the modular
transformations~\eqref{eq:Yukawas_Trafos_L}, with the matrices
\begin{equation}\label{eq:rho_Y_6_224}
\rho_{\rep{4}}(\widetilde{S})=- \,\frac{\mathrm{e}^{\I \frac{\pi}{4}}}{\sqrt{4}}
\begin{pmatrix}
 1 & 1 & 1 & 1 \\
 1 & \I & -1 & -\I \\
 1 & -1 & 1 & -1 \\
 1 & -\I & -1 & \I \\
\end{pmatrix}
\quad\text{and}\quad
\rho_{\rep{4}}(\widetilde{T})=
\begin{pmatrix}
 1 & 0 & 0 & 0 \\
 0 & \mathrm{e}^{\frac{\pi\I }{4}} & 0 & 0 \\
 0 & 0 & -1 & 0 \\
 0 & 0 & 0 & \mathrm{e}^{\frac{\pi\I }{4}} \\
\end{pmatrix}\;.
\end{equation}
Analogously to what we have done in \cref{sec:ToyModel_336}, due to the
relations of the Yukawa couplings in \cref{eq:Yukawas_fourplet}, we can
reduce the representation matrices through the projection matrix 
\begin{subequations}\label{eq:ProjectionMatrix4to3}
\begin{align}
P_{4\to 3}&=
\begin{pmatrix}
 1 & 0 & 0 \\
 0 & \frac{1}{\sqrt{2}} & 0 \\
 0 & 0 & 1 \\
 0 & \frac{1}{\sqrt{2}} & 0 \\
\end{pmatrix}\; .
\end{align}
\end{subequations}
We can define the triplet of independent Yukawa couplings through
$Y_{\rep{3}}=P_{4\to 3}^{T}Y_{\rep{4}}$, which transforms under the
representation matrices given by
\begin{subequations}\label{eq:rho_Y_6to4_224}
\begin{align}
\rho_{\rep{3}}(\widetilde{S})&= P_{4\to 3}^{T}\rho_{\rep{4}}(\widetilde{S})P_{4\to 3}
= -\,\frac{\mathrm{e}^{\I \frac{\pi}{4}}}{\sqrt{4}}
\begin{pmatrix}
 1 & \sqrt{2} & 1 \\
 \sqrt{2} & 0 & -\sqrt{2} \\
 1 & -\sqrt{2} & 1 \\
\end{pmatrix}\;, \\ 
\notag \\ 
\rho_{\rep{3}}(\widetilde{T})&= P_{6\to 4}^{T}\rho_{\rep{6}}(\widetilde{T})P_{6\to 4}
=\begin{pmatrix}
 1 & 0 & 0 \\
 0 & \mathrm{e}^{\frac{\pi\I }{4}} & 0 \\
 0 & 0 & -1 \\
\end{pmatrix}\;.
\end{align}
\end{subequations}
The transformation matrices of the doublets $L^{i}$ and $H^{j}$ are obtained by
setting $M=2$ in \cref{eq:4Dfieldtransformations} and taking its conjugate. The
resulting matrices are given by \cref{eq:rho_Y_2}. The four--plets $R^{k}$
transform through \cref{eq:4Dfieldtransformations} for $M=4$, that is
\begin{equation}\label{eq:rho_phi_4}
\rho_{\rep{4}'}^{\phi}(\widetilde{S})= \frac{\mathrm{e}^{\I \frac{\pi}{4}}}{\sqrt{4}}
\,\begin{pmatrix}
 1 & 1 & 1 & 1 \\
 1 & \I & -1 & -\I \\
 1 & -1 & 1 & -1 \\
 1 & -\I & -1 & \I \\
\end{pmatrix}
\quad\text{and}\quad
\rho_{\rep{4}'}^{\phi}(\widetilde{T})=
\begin{pmatrix}
 1 & 0 & 0 & 0 \\
 0 & \mathrm{e}^{-\frac{3 \pi\I }{4}} & 0 & 0 \\
 0 & 0 & -1 & 0 \\
 0 & 0 & 0 & \mathrm{e}^{-\frac{3 \pi\I }{4}} \\
\end{pmatrix}\;.
\end{equation}
The representation matrices in
\cref{eq:rho_Y_2,eq:rho_Y_6_224,eq:rho_Y_6to4_224,eq:rho_phi_4} fulfill the
conditions~\eqref{eq:tildeGamma_4N-presentation} for $N=2$ and
\eqref{eq:tildeGamma_4(2)-presentation} which implies that we have a theory
endowed with a $\widetilde{\Gamma}_{2 \lambda=8}\cong[768,1085324]$ metaplectic
flavor symmetry. In this model, the four--plet of the Yukawa couplings
decomposes  as $\rep{4}=\rep{3}\oplus\rep{1}'$, where the singlet vanishes. On
the other hand, the matter four--plets decompose as
$\rep{4}'=\rep{3}'\oplus\rep{1}''$ whereas the matter doublets $\rep{2}$ are
irreducible. While all these representations are unfaithful, the
combination $\rep{4}'\oplus\rep{2}$ provides us with a faithful
representation content of $\widetilde\Gamma_{8}$. The tensor products of
$\rep{4}'$ and $\rep{2}$ produce all $\widetilde\Gamma_8$ representations.

\subsubsection{Model with $\mathcal{I}_{ab}=1$,  $\mathcal{I}_{ca}=2$ and $\mathcal{I}_{bc}=-3$}
\label{sec:ToyModel_123}

It is important to show that our conjecture \eqref{eq:conjecture} that we have a
$\widetilde{\Gamma}_{2 \lambda}$ invariant superpotential is not only valid for
repeated values of $\mathcal{I}_{\alpha\beta}$. To this end we consider a toy
model with the Yukawa couplings as in \cref{eq:Superpotential336Model} and with
gauge group breaking $\U3\rightarrow \U1_a\times\U1_b\times\U1_c$  by the fluxes
\begin{align}
	F &= \frac{\pi \I}{\im\tau}
	\begin{pmatrix} 
	-1 & 0 & 0 \\
    0 & -2 & 0 \\
    0 & 0 & 1 \\
    \end{pmatrix}\;.
\end{align}
Out of $3\cdot2\cdot1$ \emph{a priori} possible Yukawa couplings, $\lambda/2+1=4$ are
distinct
\begin{subequations}
\begin{align}
 Y_0&:=Y_{000}\;, & Y_1&:=Y_{011}=Y_{5}:=Y_{012}\;,
 \\
 Y_3&:=Y_{010}\;, & Y_2&:=Y_{001}=Y_{4}:=Y_{002}\;,
\end{align}
\end{subequations}
where $i$ can only be $0$, and $j$ and $k$ are understood to be modulo $2$ and
$3$ respectively. As $d=1$, this setting has a comparatively large number of
distinct couplings, i.e.\ $4$ out of $6$ entries are distinct whereas e.g.\ in
the model of \cref{sec:ToyModel_336} only $4$ out of $54$ \emph{a priori}
contractions have nontrivial distinct coefficients. In this case,
$\mathcal{I}_{ab}\ne\mathcal{I}_{ca}$,  so there is no exchange symmetry of the
type \eqref{eq:exchangesymmetries}, yet the number of independent Yukawa
couplings gets reduced due to the symmetry $Y_{0jk}=Y_{0j,3-k}$. Unlike the
transformation that ensured the equality of Yukawa entries in the model
discussed in \cref{sec:ToyModel_336}, this symmetry is not a nontrivial outer
automorphism of the 4D continuous gauge symmetries. The six--plet of
holomorphic Yukawa couplings $\mathcal Y_{\widehat{\alpha}}$ transform with  the
matrices from \cref{eq:rho_Y_6} that can be reduced by using 
\cref{eq:ProjectionMatrix6to4} to a four--plet, which then transforms according
to \cref{eq:rho_Y_6to4}. Furthermore, using \cref{eq:4Dfieldtransformations}, we
see that the singlet $\phi^{0,\mathcal{I}_{ab}=1}$ is invariant under modular
transformations, the doublet $\phi^{j,\mathcal{I}_{ca}=2}$ transforms with the
representation matrices from \cref{eq:rho_Y_2}, and the triplet
$\phi^{k,\mathcal{I}_{cb}=3}$ transforms using \cref{eq:rho_phi_3}. It can be
shown that all these matrices satisfy the
conditions~\eqref{eq:tildeGamma_4N-presentation} for $N=3$ and
\cref{eq:tildeGamma_4(3)-presentation}. Therefore, the superpotential is
invariant under the finite metaplectic flavor symmetry  $\widetilde{\Gamma}_{2
\lambda=12}$ of order 2304.

\subsection{Comments on the relation to bottom--up constructions}

As we have seen, the models derived from explicit tori give rise to the finite
metaplectic groups, which have been discussed e.g.\ in \cite{Liu:2020msy} in
the context of bottom--up model building. The models presented here do not
attempt to make an immediate connection to particle phenomenology. At first
sight, it seems to be hard to derive the models of \cite{Liu:2020msy} from tori
for at least two reasons:
\begin{enumerate}
 \item our fields all have modular weight $-\nicefrac{1}{2}$ while in
  \cite{Liu:2020msy} they come with a variety of weights, and
 \item we have additional symmetries like the outer automorphism symmetry
  \eqref{eq:exchangesymmetries} and residual gauge factors.
\end{enumerate}
On the other hand, deriving the modular transformations from an explicit
higher--dimensional model has the virtue that normalization of the fields is
known at tree level, and that otherwise free parameters get fixed. Of course,
the K\"ahler potential is not exact, apart from the usual 4D corrections there
are additional terms contributing (cf.\ \cite{DiVecchia:2008tm}), yet the point
that there is a zeroth order classical form plus corrections, which are under
control. On the other hand, in the \ac{SB} approach every invariant
K\"ahler potential is as good as others \cite{Chen:2019ewa}, and there are thus
large uncertainties. An additional benefit of deriving the modular flavor
symmetries from explicit tori is the geometrical intuition one can develop for
the Yukawa couplings, cf.\ our discussion at the end of \cref{sec:ToyModel_336}.

One may now wonder if the price that one has to pay for all these benefits is
the inability to construct semi--realistic models. In what follows, we will
argue that this is not the case. First of all, the $\mathds{T}^2$ model is just
a building block of a more complete story. As explained in
\cite{Cremades:2004wa}, these models are dual to some intersecting $D$--brane
constructions. Moreover, the couplings of the latter are closely related to
heterotic string compactifications \cite{Abel:2003yx}, which provide us with a
large number of potentially realistic models~\cite{Ibanez:2012zz}.\footnote{We 
adopt the convention to call models with realistic and unrealistic features 
``semi--realistic'' while ``potentially realistic'' models are constructions that 
have no obviously unrealistic features, but have not yet been analyzed in enough 
detail to be called realistic.} These more complete settings come with a variety of
modular weights \cite{Ibanez:1992hc}. Second, even if one is not adding more
dimensions to the construction, fields with higher modular weights can emerge as
composites of fields with modular weight $-\nicefrac{1}{2}$. That is, if
``quarks'' of a model with an $\SU{N_c}$ have modular weights
$-\nicefrac{1}{2}$, then the ``baryons'' will have weights $-\nicefrac{N_c}{2}$.

\section{Comments on the role of supersymmetry}
\label{sec:CommentsOnSUSY}

Let us comment on the role \ac{SUSY} plays in the discussion. While  Cremades et
al.\ work in a supersymmetric theory, they mention \cite[see the beginning of
section~5.3]{Cremades:2004wa} that their derivation ``in principle is valid for
toroidal compactifications where supersymmetry might be broken explicitly''. Of
course, if one wants to claim that the couplings that one has computed are
Yukawa couplings, one needs to make sure that one computes the overlap between
two fermionic and one bosonic zero--modes. In supersymmetric models there is no
problem because the superpartners are described by the same wave functions. 

In a model without low--energy \ac{SUSY} one may be worried that quantum
corrections lead to uncontrollable corrections to the wave function of the
scalar. This is generally a very valid concern, yet is as recently been observed
that in the magnetized tori there is an interesting cancellation of corrections
to the scalar mass
\cite{Buchmuller:2016gib,Ghilencea:2017jmh,Buchmuller:2018eog,Hirose:2019ywp}.
While this has not yet led to a complete solution of the hierarchy problem in
the \ac{SM}, it does suggest that in the context of the very models that we were
led to consider for the sake of deriving modular flavor symmetries the
situation is ``better'' than in other nonsupersymmetric completions of the
\ac{SM} with a high \ac{UV} scale. In fact, similar cancellations have been
reported in \cite{Dienes:1994np}, where they were attributed to modular
symmetries.

In a bit more detail, one could imagine a torus compactification
in which the Yukawa couplings emerge as outlined in \cite{Cremades:2004wa},
namely as the overlap of three wave functions. These wave functions describe two
fermions and one scalar, such as the \ac{SM} Higgs. If the scalars remain light,
they will still be approximate zero--modes, and thus the profile is
approximately given by \cref{eq:ZeroModesWavefunctions}. So the Yukawa couplings
will, to some good approximation, be the ones of \cref{eq:Yijk}. So
supersymmetry is not instrumental for having models with modular flavor
symmetries.

\section{Summary}
\label{sec:Summary}

We discussed how modular flavor symmetries derive from explicit tori which
are endowed with magnetic fluxes. Using Euler's Theorem, we have derived a
closed--form expression for the Yukawa couplings between zero modes.
This expression generalizes the results of the literature to arbitrary flux
parameters $\mathcal{I}_{ab}$ and $\mathcal{I}_{ca}$, which fix
$\mathcal{I}_{bc}$, and is not restricted to the special case in which one flux
parameter equals $1$. Each entry of the Yukawa tensor is a single
$\vartheta$--function, i.e.\ the holomorphic Yukawa couplings take the form
\begin{equation}\label{eq:HolomorphicYukawaSummary}
 \mathcal{Y}_{ijk}(\tau)=
 \vartheta\Orb{\widehat{\alpha}_{ijk}/\lambda}{0}(0,\lambda\,\tau)\;,
\end{equation}
where 
\begin{equation}\label{eq:alphaijk_def}
 \widehat{\alpha}_{ijk}=
 \mathcal{I}_{ca}'\,i-\mathcal{I}_{ab}'\,j
 +\mathcal{I}_{ca}'\,\left(\mathcal{I}_{ab}'\right)^{\phi(|\mathcal{I}_{bc}'|)}
 \,(k-i-j)\mod\lambda\;.
\end{equation}
Here, $\phi$ denotes the Euler $\phi$--function,
$\mathcal{I}_{\alpha\beta}'=\mathcal{I}_{\alpha\beta}/d$ for
$\alpha,\beta\in\{a,b,c\}$, 
$d=\gcd\bigl(|\mathcal{I}_{ab}|,|\mathcal{I}_{ca}|,|\mathcal{I}_{bc}|\bigr)$,
$\lambda=\lcm\bigl(|\mathcal{I}_{ab}|,|\mathcal{I}_{ca}|,|\mathcal{I}_{bc}|\bigr)$,
and $\tau$ denotes the half--period ratio of the torus. The condensed form for
holomorphic Yukawa couplings as single $\vartheta$--functions is instrumental
for deriving the symmetries between Yukawa couplings. As we have seen, these
symmetries include outer automorphisms of the low--energy gauge symmetry. These
couplings are modular forms of weight $\nicefrac{1}{2}$ of level $2\lambda$ that
build representations under the metaplectic modular flavor symmetry
$\widetilde{\Gamma}_{2\lambda}$. There are at most $1+\lambda/2$ distinct Yukawa
couplings, which transform as a (generically reducible) representation
$\rep\lambda$ of $\widetilde{\Gamma}_{2\lambda}$. This means that e.g.\ a model
with flux parameters 
$(\mathcal{I}_{ab},\mathcal{I}_{ca},\mathcal{I}_{bc})=(1,2,-3)$ has as many
independent Yukawa couplings as a model with fluxes $(3,3,-6)$.  We have
commented on the geometric interpretation of the Yukawa couplings, and that the
$\widehat{\alpha}_{ijk}$ in \cref{eq:HolomorphicYukawaSummary} corresponds to a
separation of the states, and lead to an exponential suppression with an
exponent $\min(\widehat{\alpha}_{ijk},1-\widehat{\alpha}_{ijk})\,\im\tau$ for
unsuppressed $\im\tau$. The 4D fields have a well--defined transformation
behavior under  $\widetilde{\Gamma}_{2\lambda}$, regardless of whether the flux
is even or odd, and have weight $k_\phi = -\nicefrac{1}{2}$. Our analysis is
restricted to a magnetized torus $\mathds{T}^2$ and its half--period ratio
$\tau$, which is contained in the so--called complex structure modulus. We also
set the Wilson lines to zero, and largely disregarded the K\"ahler modulus.
While this is sufficient to derive meaningful modular flavor symmetries, this
analysis may be thought of as a building block of more complete, perhaps stringy
models. It will be interesting to derive a generalization of
\cref{eq:HolomorphicYukawaSummary} for such constructions.

We have also commented on the role that supersymmetry plays in these
constructions. As already pointed out in \cite{Cremades:2004wa}, supersymmetry
is not instrumental as long as the profiles of the zero--modes do not get
distorted too much. More recent
analyses~\cite{Buchmuller:2016gib,Ghilencea:2017jmh,Buchmuller:2018eog,Hirose:2019ywp}
indicate that magnetized tori have certain unusual properties in that scalar
masses seem to be immune to quantum corrections even without supersymmetry. This
means that one can plausibly disentangle modular flavor symmetries from the
question of low--energy supersymmetry.

\subsection*{Acknowledgments}

The work of Y.A.\ was supported by Kuwait University. The work of M.-C.C., M.R.\ and S.S.\ was supported by the National Science
Foundation, under Grant No.\ PHY-1915005.  The work of S.R.-S.\ was
partly supported by CONACyT grant F-252167. This work is also supported by
UC-MEXUS-CONACyT grant No.\ CN-20-38.

\appendix

\section[Theta-functions]{$\vartheta$--functions}
\label{sec:theta-functions}

In this appendix, we collect some relevant facts on the $\vartheta$--functions.
Our conventions are based on \cite{Mumford:Theta1} and \cite{Polchinski:1998rq}.
One defines 
\begin{equation}\label{eq:DefEllipticThetaAlphaBeta}
 \vartheta\orb{\alpha}{\beta}(z,\tau)
 :=
 S_\beta\,T_\alpha\,\vartheta(z,\tau)
 =
 \mathrm{e}^{2\pi\I\,\alpha\,\beta}
 \,T_\alpha\,S_\beta\,\vartheta(z,\tau)\;,
\end{equation}
where \cite[p.~4]{Mumford:Theta1}
\begin{equation}\label{eq:DefineBasicTheta}
 \vartheta(z,\tau):=\sum\limits_{\ell\in\mathds{Z}}
 \exp(\I\,\pi\,\ell^2\,\tau)\,\exp(2\pi\I\,\ell\,z)
\end{equation}
with $\im\tau>0$, and \cite[p.~6]{Mumford:Theta1}
\begin{subequations}
\begin{eqnarray}
 (S_\beta f)(z) 
 & := & f(z+\beta)\;,\label{eq:Sbeta}\\
 (T_\alpha f)(z) 
 & := & 
 \mathrm{e}^{\I\,\pi\,\alpha^2\,\tau+2\pi\I\,\alpha\,z}\,f(z+\alpha\,\tau)
 \;.
\end{eqnarray}
\end{subequations}
This immediately gives us (cf.\ \cite[p.~214~f.]{Polchinski:1998rq})
\begin{equation}\label{eq:ThetaSum1}
 \vartheta\orb{\alpha}{\beta}(z,\tau)
 =
 \sum\limits_{\ell=-\infty}^\infty
 \mathrm{e}^{\I\pi\,(\alpha+\ell)^2\,\tau}
 \,\mathrm{e}^{2\pi\I\,(\alpha+\ell)\,(z+\beta)}\;.
\end{equation}
For an integer $n\in\mathds{Z}$ one has torus periodicities
\begin{subequations}
\begin{align}
 \vartheta\orb{\alpha}{\beta}(z+n,\tau)
 & =  
 \mathrm{e}^{2\pi\I\,n\,\alpha}\,
 \vartheta\orb{\alpha}{\beta}(z,\tau)
 \;,\label{eq:z+n}\\
 \vartheta\orb{\alpha}{\beta}(z+n\,\tau,\tau)
 & =
 \mathrm{e}^{-\I\pi\,n^2\,\tau-2\pi\I\,n\,(z+\beta)}\,
 \vartheta\orb{\alpha}{\beta}(z,\tau)
 \;.\label{eq:z+ntau}  
\end{align}
\end{subequations}
Further, the $\vartheta$--function have several periodicities in the
characteristics $\alpha$ and $\beta$,
\begin{subequations}
\begin{align}
 \vartheta\orb{\alpha+1}{\beta}(z,\tau)
 &  =
 \vartheta\orb{\alpha}{\beta}(z,\tau)
 \;,\label{eq:ShiftAlpha}\\
 \vartheta\orb{\alpha}{\beta+1}(z,\tau)
 &  =
 \mathrm{e}^{2\pi\I\, \alpha}\,
 \vartheta\orb{\alpha}{\beta}(z,\tau)
 \;.\label{eq:ShiftBeta}
\end{align}
\end{subequations}
The behavior under modular transformation is
\begin{subequations}\label{eq:ThetaModular}
\begin{align}
 \vartheta\orb{\alpha}{\beta}(z,\tau+1)
 &= 
 \mathrm{e}^{-\I\,\pi\,\alpha\,(\alpha+1)}\, 
 \vartheta\orb{\alpha}{\beta+\alpha+\frac{1}{2}}(z,\tau)
 \;,\label{eq:ThetaModularT}\\
\vartheta\orb{\alpha}{\beta}\left(-\frac{z}{\tau},-\frac{1}{\tau}\right)
 &= \sqrt{-\I\,\tau}\,
 \mathrm{e}^{2\pi\I\,\left(\frac{z^2}{2\tau}+\alpha\,\beta\right)}\,
 \vartheta\orb{-\beta}{\alpha}(z,\tau)\;.\label{eq:ThetaModularS}
\end{align}
\end{subequations}
Another useful formula is
\begin{align}
 \vartheta\orb{0}{\nicefrac{j}{M}}\left(z,\frac{\tau}{M}\right)
  &=
  \sum_{k=0}^{M-1}e^{2\pi\text{i}jk/M}\vartheta\begin{bmatrix}\frac{k}{M} \\ 0 \end{bmatrix}(Mz,M\tau)
  \;.\label{eq:Poisson}
\end{align}

\section{Torus integration}
\label{sec:Normalization}

The torus is defined by two lattice vectors, which can be chosen as $e_{1}=2\pi
R$  and $e_{2} = 2\pi R\,\tau$, where the real, dimensionful quantity $R$ sets
the length of one  lattice vector, and $\tau$ with $\im\tau>0$ is  the
half--period ratio. In this basis, the torus metric reads
\begin{equation}\label{eq:torusmetric}
G=(2\pi R)^{2}\begin{pmatrix}
1  & \re \tau \\
\re\tau & |\tau |^{2}
\end{pmatrix}\;.
\end{equation}
We can define the \emph{fundamental domain} of the torus as
\begin{equation}\label{eq:FundamentalDomainF}
 \mathds{T}^{2}=\{{z}\in\mathds{C}\;;~~ z=x\,e_{1}+y\,e_{2}
 \ \text{ with }\ 0\le x,y\le 1\}\;.
\end{equation}
It is straightforward to verify that the Jacobian of the transformation $(\re
z,\im z)\mapsto(x,y)$ is given by $(2\pi R)^2\im\tau$. Therefore, the integrals
of an arbitrary function $f(z)$ over the fundamental domain are given by
\begin{equation}\label{eq:Dictionary_z_vs_x_and_y_integral}
 \int\limits_{\mathbb{T}^{2}}\!\D^2z\,f(z)
   = (2\pi R)^{2}\im\tau\,\int\limits_{0}^{1}\!\D x\,\int\limits_{0}^{1}\!\D y\,f(xe_1+ye_2)\;.
\end{equation}
Let us now look at constant modes on the torus, or, equivalently integrate over
the torus $\mathds{T}^2$ to determine its volume. We then have
\begin{equation}
 \operatorname{vol}(\text{torus})=
 \int\limits_{\mathds{T}^{2}}\!\D^2z\,1=(2\pi R)^2\im\tau =:\mathcal A\;.
\end{equation}

\section{Explicit verification of the boundary conditions for transformed wave
functions}
\label{app:boundary_conditions_explicit_check}

\subsection{$S$ transformation}
\label{app:boundary_conditions_explicit_check_S}

We now compute the $S$ transformation, $\tau\mapsto-1/\tau$ 
(cf.~\cref{eq:Ttransformation}), of \cref{eq:ZeroModesWavefunctions}. We have
\begin{align}
\MoveEqLeft \wavefunction \left(-\frac{z}{\tau},-\frac{1}{\tau},0\right)=\bigg(\frac{2M\im\frac{-1}{\tau}}{{\mathcal{A}}^{2}}\bigg)^{1/4}\exp\bigg(\I \pi M \frac{z}{\tau}\frac{\im\frac{z}{\tau}}{\im\frac{-1}{\tau}}\bigg)
 \,\vartheta\orb{\frac{j}{M}}{0}\bigg(-\frac{Mz}{\tau},-\frac{M}{\tau}\bigg)\nonumber\\
 &=\frac{\mathcal{N}}{\sqrt{|\tau |}}\exp\bigg(\I\pi M \frac{z}{\tau}\frac{\im{z\bar{\tau}}}{\im{\tau}}\bigg)\sqrt{-\I\frac{\tau}{M}}\mathrm{e}^{2\pi M\I\frac{z^{2}}{2\tau}}
	\,\vartheta\orb{0}{\frac{j}{M}}\bigg(z,\frac{\tau}{M}\bigg)\nonumber\\
 &=\frac{\mathcal{N}}{\sqrt{|\tau |}}\exp\bigg(\I\pi M \frac{z}{\tau}\frac{\im{z\bar{\tau}}}{\im{\tau}}\bigg)\sqrt{-\I \frac{\tau}{M}}\mathrm{e}^{\I  \pi M \frac{z^{2}}{\tau}}
	\sum_{k=0}^{M-1}\mathrm{e}^{2\pi\I\, jk/M}
	\,\vartheta\orb{\frac{k}{M}}{0}(Mz,M\tau)\nonumber\\
 &=\frac{\mathrm{e}^{{\frac{\I \pi}{4}}}}{\sqrt{M}}\bigg({-}\frac{\tau}{|\tau|}\bigg)^{1/2}\sum_{k=0}^{M-1}\mathrm{e}^{\frac{2\pi\I jk}{M}}\bigg[\mathcal{N}\exp\bigg(\frac{\I \pi M z}{\tau}{\frac{\im z\bar{\tau}}{\im\tau}}+\I M\pi\frac{z^{2}}{\tau}\bigg)
	 \,\vartheta\orb{\frac{k}{M}}{0}(Mz,M\tau)\bigg]\nonumber\\
 &=\frac{\mathrm{e}^{{\I} \frac{\pi}{4}}}{\sqrt{M}}\bigg({-}\frac{\tau}{|\tau |}\bigg)^{1/2}\sum_{k=0}^{M-1}\mathrm{e}^{\frac{2\pi \I jk}{M}}\mathcal{N}\mathrm{e}^{i\pi M z\frac{\im z}{\im \tau}}
    \,\vartheta\orb{\frac{j}{M}}{0}(Mz,M\tau)\;,
    \label{eq:Stransform1}
\end{align}
where we used \eqref{eq:Normalization}, \eqref{eq:ThetaModularS} and
\eqref{eq:Poisson} in the first, second and third lines respectively. We thus
arrive at  \eqref{eq:StransformWavefunction}. Therefore, the $S$ modular
transformation of the zero modes is
\begin{align}
 \wavefunction\left(-\frac{z}{\tau},-\frac{1}{\tau},0\right)
 &=\frac{\mathrm{e}^{{\I} \frac{\pi}{4}}}{\sqrt{M}}
 \bigg({-}\frac{\tau}{|\tau |}\bigg)^{1/2}
 \sum_{k=0}^{M-1}\mathrm{e}^{2\pi \I jk/M}\,\psi^{k,M}(z,\tau,0)\;.
    \label{eq:StransformWavefunction}
\end{align}
It is straightforward to see that the wave function of
\cref{eq:StransformWavefunction} satisfies the boundary conditions of
\cref{eq:boundary1STransformed,eq:boundary2STransformed}. Note that
\begin{multline}
 \wavefunction \bigg( -\frac{z}{\tau} + 1, -\frac{1}{\tau},0 \bigg) 
 = \wavefunction \bigg( -\frac{(z-\tau )}{\tau} , -\frac{1}{\tau},0 \bigg)
 \\
 =\frac{\mathrm{e}^{{\I } \frac{\pi}{4}}}{\sqrt{M}}\bigg({-}\frac{\tau}{|\tau
 |}\bigg)^{1/2}\sum_{k=0}^{M-1}\mathrm{e}^{\frac{2\pi \I jk}{M}}\,
 \psi^{k,M}(z-\tau,\tau,0) 
 = \exp\bigg(-\I \pi M \frac{\im z \bar{\tau}}{\im \tau}   \bigg)\, 
\wavefunction \bigg(-\frac{z}{\tau},-\frac{1}{\tau} , 0 \bigg) 
\label{eq:SatisfiesS1}
\end{multline}
and
\begin{multline}
\wavefunction \bigg( -\frac{z}{\tau} - \frac{1}{\tau}, -\frac{1}{\tau }, 0 \bigg) 
= \wavefunction \bigg(  -\frac{(z+1)}{\tau}, -\frac{1}{\tau }, 0 \bigg)\\
= \frac{\mathrm{e}^{{\I } \frac{\pi}{4}}}{\sqrt{M}}\bigg({-}\frac{\tau}{|\tau |}\bigg)^{1/2}\sum_{k=0}^{M-1}\mathrm{e}^{\frac{2\pi \I jk}{M}}
\wavefunction[k,M](z+1,\tau,0)
= \exp\bigg(\frac{\I \pi M \im z }{\im \tau}\bigg)\wavefunction \bigg(-\frac{z}{\tau},-\frac{1}{\tau}, 0\bigg)
\;.\label{eq:SatisfiesS2}
\end{multline}
Thus, from \cref{eq:SatisfiesS1,eq:SatisfiesS2} we can see that the $S$
transformed zero mode follows the boundary conditions of
\cref{eq:boundary1STransformed,eq:boundary2STransformed} for both odd and even
$M$. 

\subsection{$T$ transformation}
\label{app:boundary_conditions_explicit_check_T}

Now, we compute the transformed wave function $\wavefunction (z,\tau + 1 ,0)$
and check that it satisfies both \cref{eq:boundary1TTransformed} and
\cref{eq:boundary2TTransformed}. Applying the $T$ modular transformation in
\cref{eq:ZeroModesWavefunctions} gives
\begin{align}
    \wavefunction (z,\tau + 1,0)
	&=
	\mathcal{N}\mathrm{e}^{\I\pi\, M z \frac{\im
	z}{\im\tau}}\vartheta\orb{\frac{j}{M}}{0}
	\bigl(Mz,M(\tau+1),0\bigr) \nonumber\\
    &=\mathrm{e}^{-\I\,\pi j\bigl(\frac{j}{M}+1\bigr)}\,\mathcal{N}
	\mathrm{e}^{\I\,\pi M z \frac{\im z}{\im\tau}}\,
	\vartheta\orb{\frac{j}{M}}{j+\frac{M}{2}}(Mz,M\tau) \nonumber \\
    &=\mathrm{e}^{-\I\,\pi j\bigl(\frac{j}{M}+1\bigr)}\,
	\mathcal{N}\mathrm{e}^{\I\,\pi M z \frac{\im z}{\im\tau}}
	\,\mathrm{e}^{2\pi\I\,\frac{j}{M}j}\vartheta\orb{\frac{j}{M}}{\frac{M}{2}}(Mz,M\tau)\nonumber\\
    &=\mathrm{e}^{-\I\,\pi j\bigl(1-\frac{j}{M}\bigr)}\,
	\mathcal{N}\mathrm{e}^{\I\,\pi M z \frac{\im z}{\im\tau}}
	\,\vartheta\orb{\frac{j}{M}}{\frac{M}{2}}(Mz,M\tau)\;,
    \label{eq:transformedtau+1}
\end{align}
where we used \eqref{eq:ThetaModularT} and \eqref{eq:ShiftBeta} in the second
and third line, respectively. Defining
$\widetilde{\mathcal{N}}:=\mathrm{e}^{-\I\,\pi
j\bigl(1-\frac{j}{M}\bigr)}\mathcal{N}$ we can thus write
\begin{equation}
    \wavefunction (z,\tau+1,0)=\widetilde{\mathcal{N}} \mathrm{e}^{\I\,\pi M z \frac{\im z}{\im\tau}}
	\vartheta\orb{\frac{j}{M}}{\frac{M}{2}}(Mz,M\tau)\;.
    \label{eq:WavefunctionTtransformed}
\end{equation}
Now we check that \cref{eq:WavefunctionTtransformed} satisfies boundary
conditions given by \cref{eq:boundary1TTransformed,eq:boundary2TTransformed}.
The first boundary condition is satisfied as shifting $z\rightarrow z + 1$ in
\cref{eq:WavefunctionTtransformed} gives
\begin{align}
\wavefunction (z+1,\tau + 1,0 )~&=~\widetilde{\mathcal{N}} \mathrm{e}^{\I\,\pi M (z+1) \frac{\im z}{\im\tau}}
	\vartheta\orb{\frac{j}{M}}{\frac{M}{2}}(M(z+1),M\tau)\notag\\
    &=~\exp\bigg(\I\,\frac{\pi M\im z}{\im\tau}\bigg)\,\widetilde{\mathcal{N}} 
	\mathrm{e}^{\I\,\pi M (z+1) \frac{\im z}{\im\tau}}
	\mathrm{e}^{2\pi\I\, M \frac{j}{M}}
	\vartheta\orb{\frac{j}{M}}{\frac{M}{2}}(Mz,M\tau)\notag\\
	&=~\exp\bigg(\I\,\frac{\pi M\im z}{\im\tau}\bigg)
	\wavefunction (z,\tau + 1,0 )\;,
\label{eq:Satisfies1}
\end{align}
where we used \cref{eq:z+n} in the second line. For
\cref{eq:boundary2TTransformed} we have
\begin{align}
\MoveEqLeft\wavefunction (z+\tau+1,\tau + 1,0)
	=\widetilde{\mathcal{N}}\mathrm{e}^{\frac{\I\,\pi M
	 (z+\tau+1)}{\im\tau}\im(z+\tau+1)}\vartheta\orb{\frac{j}{M}}{\frac{M}{2}}(M(z+\tau+1),M\tau )\nonumber\\
   &=\widetilde{\mathcal{N}}\mathrm{e}^{\frac{\I\,\pi M}{\im\tau}[z\im z+(\tau+1)\im z+z\im(\tau + 1)+(\tau + 1)\im(\tau + 1)}
   \,\mathrm{e}^{2\pi\I\,\frac{j}{M}M}
	\vartheta\orb{\frac{j}{M}}{\frac{M}{2}}(M(z+\tau),M\tau )\nonumber\\
    &=\widetilde{\mathcal{N}}\mathrm{e}^{\frac{\I\,\pi M}{\im\tau}[z\im z+(\tau+1)\im z+z\im(\tau + 1)+(\tau + 1)\im(\tau + 1)}
	\,
	\mathrm{e}^{-\pi\I\,\tau M-2\pi\I\,(Mz+{\frac{M}{2})}}
	\vartheta\orb{\frac{j}{M}}{\frac{M}{2}}(Mz,M\tau )\nonumber\\
    &=\mathrm{e}^{\frac{\I\,\pi M}{\im\tau}[(\tau+1)\im z+z\im(\tau+1)+(\tau+1)\im(\tau+1)-\tau\im\tau-2z\im\tau-\im\tau]}
	\,
	\widetilde{\mathcal{N}}\mathrm{e}^{\frac{\I\,\pi M}{\im\tau}z\im z}
	\vartheta\orb{\frac{j}{M}}{\frac{M}{2}}(Mz,M\tau )\nonumber\\
    &=\exp\bigg(\frac{\I\,\pi M}{\im\tau}\im(\bar{\tau}+1)z\bigg)\,
	\wavefunction (z,\tau+1,0)
	\;,
    \label{eq:Satisfies2}
\end{align}
where we have used \cref{eq:z+n,eq:z+ntau} in the second and third line
respectively.  Therefore, the transformed modular wave function given by 
\cref{eq:WavefunctionTtransformed} follows the transformed boundary conditions
of \cref{eq:boundary1TTransformed,eq:boundary2TTransformed} for even and odd
$M$. 

Let us now tackle the problem of expressing the $T$ transformed wave functions
in terms of the original ones. As noted in section
\eqref{sec:ModularTransformationsOfWaveFunctions}, for odd values of $M$ it is
not possible to express the $T$  transformed wave functions in terms of the
original ones because of \eqref{eq:wrongboundary}.  At the level of the wave
functions, one can refer to \cref{eq:WavefunctionTtransformed} and see that if
$M$ is even, then using  \eqref{eq:ShiftBeta} confirms
\eqref{eq:Kikuchi:2020frp-37}. However, if $M$ is odd, in order to make use of
\eqref{eq:ShiftBeta} we need to shift the $z$ coordinate as $z \mapsto z +
\Delta z$  with real $\Delta z$. Using \cref{eq:WavefunctionTtransformed} we
have
\begin{align}\label{eq:bshiftdetailed}
\MoveEqLeft\wavefunction(z + \Delta z,\tau+1,0)=\widetilde{\mathcal{N}} \mathrm{e}^{\I\,\pi M (z+\Delta z) \frac{\im z}{\im\tau}}
	\vartheta\orb{\frac{j}{M}}{\frac{M}{2}}(M(z+\Delta z),M\tau) \notag\\
	&=\widetilde{\mathcal{N}} \mathrm{e}^{\I\,\pi M \Delta z \frac{\im z}{\im\tau}} \mathrm{e}^{\I\,\pi M z \frac{\im z}{\im\tau}}
	\vartheta\orb{\frac{j}{M}}{0}(M(z+\Delta z+\tfrac{1}{2} ) ,M\tau),
\end{align}
where we have used \eqref{eq:ThetaSum1} to rewrite the lower characteristic of
the $\vartheta$--function as a shift in the $z$ coordinate. Therefore, if we
assume that $\Delta z$ is a half--integer number, we can use \eqref{eq:z+n}
which gives
\begin{align}\label{eq:bshiftdetailed2}
\MoveEqLeft\wavefunction(z + \Delta z,\tau+1,0)=\widetilde{\mathcal{N}}\mathrm{e}^{\I\,\pi M \Delta z \frac{\im z}{\im\tau}} \mathrm{e}^{\I\,\pi M z \frac{\im z}{\im\tau}}\mathrm{e}^{2\pi\I j (\frac{1}{2}+\Delta z)}
	\vartheta\orb{\frac{j}{M}}{0}(M z,M\tau)\notag\\
	&=\mathrm{e}^{\I\,\pi M \Delta z \frac{\im z}{\im\tau}}\mathrm{e}^{\pi\I j \left( \frac{j}{M}+2\Delta z\right)}\wavefunction (z, \tau, 0).
\end{align} 
Note that in order to use \eqref{eq:z+n},  $M\left( \frac{1}{2}+\Delta z\right)$
needs to be an integer. Therefore, for odd $M$, $\Delta z$ needs to be
half--integer, whereas for even $M$ both integer and half--integer $\Delta z$
are valid choices. After a redefinition of $z \rightarrow z - \Delta z$ with
some half--integer $\Delta z$,
\begin{align}
	\wavefunction(z,\tau,0)\xmapsto{~T~}
	\mathrm{e}^{\I\pi M \Delta z \frac{\im z}{\im\tau}}
	\mathrm{e}^{\I\pi\frac{j^2}{|M|}+2\I\pi j \Delta z}\,
	\wavefunction(z- \Delta z,\tau,0)\;,
\end{align}
and this is valid for both even and odd values of $M$.

\section{Symmetries between the Yukawa couplings}
\label{app:IndependentYukawas}

In this appendix we identify additional relations between the Yukawa couplings
given in \cref{eq:explicit_Yukawa}. Yukawa entries with different $i$, $j$
and/or $k$ are equal if the upper characteristic,
\begin{equation}
 \frac{\mathcal{I}_{ca}'\,i-\mathcal{I}_{ab}'\,j +\mathcal{I}_{ca}'\,\left(\mathcal{I}_{ab}'\right)^{\phi\left(|\mathcal{I}_{bc}'|\right)}\,
			(k-i-j)}{\lambda}=:u_{ijk}\;,
\end{equation}
with $\mathcal I_{\alpha\beta}'=\mathcal I_{\alpha\beta}/d$,
is the same. For instance, suppose $i' = i +r$, $j' = j +s$ and $k' = k + r +
s$, so that $i',\,j'$ and $k'$ also satisfy the selection rule (cf.\ 
\cref{eq:ijk_selection_rule_1}). Then, for values of $r$ and $s$ satisfying
\begin{align}\label{eq:relrs}
	\mathcal{I}_{ca}\,r-\mathcal{I}_{ab}\,s= 0\;, 
\end{align}
we find that $u_{ijk} = u_{i'\,j'\,k'}$, thus implying that $Y_{ijk} =
Y_{i'\,j'\,k'}$. We know that $\mathcal{I}_{ca}$ and $\mathcal{I}_{ab}$ are
divisible by
$d=\gcd\bigl(|\mathcal{I}_{ab}|,|\mathcal{I}_{ca}|,|\mathcal{I}_{bc}|\bigr)$
(cf.\ \cref{eq:d=gcd}), $u_{ijk}$ has the form
\begin{equation}
 u_{ijk}=\frac{\widehat{\alpha}_{ijk}}{\lambda}
\end{equation}
with some integer $\widehat{\alpha}_{ijk}$ given by \cref{eq:alphaijk_def}.
Further, as a shift of $u_{ijk}$ by $1$ leaves the $\vartheta$--function in the
Yukawa entry invariant (cf.\ \cref{eq:ShiftAlpha}),  there are at most
${\lambda}$ distinct entries, i.e.\ 
\begin{equation}
 u_{ijk}\in\{0,1/ {\lambda},\dots ( {\lambda}-1)/ {\lambda}\}\;.
\end{equation}
Additionally, for vanishing Wilson lines, the $\vartheta$--function takes the
simple form (cf.\ \cref{eq:ThetaSum1})
\begin{equation}\label{eq:L_Yukawas_theta}
 \vartheta\Orb{\frac{\widehat{\alpha}_{ijk}}{ {\lambda}}}{0}(0, {\lambda}\,\tau)
 =
 \sum\limits_{\ell=-\infty}^\infty
 \mathrm{e}^{\I\pi\,(\widehat{\alpha}_{ijk}/ \lambda+\ell)^2\, {\lambda} \tau}\;,
\end{equation}
which shows that 
\begin{equation}
 \vartheta\Orb{-\frac{\widehat{\alpha}_{ijk}}{ {\lambda}}}{0}(0, {\lambda}\,\tau)
 =\vartheta\Orb{\frac{\widehat{\alpha}_{ijk}}{ {\lambda}}}{0}(0, {\lambda}\,\tau)\;.
\end{equation}
Therefore, there are $ {\lambda}/2-1$ additional relations between the Yukawa
couplings, and we have at most ${\lambda}/2+1$ distinct entries. These
additional relations can manifest themselves in different ways. For instance, if
$\mathcal{I}_{ab}=\mathcal{I}_{ca}$, the overlap integral \eqref{eq:Yijk}
becomes
\begin{equation}
	Y_{ijk}=g\,\sigma_{abc}\,\int\limits_{\mathds{T}^2}\!\D^2 z\,
	\wavefunction[i,\mathcal{I}_{ab}](z,\tau,\zeta)\,
	\wavefunction[j,\mathcal{I}_{ab}](z,\tau,\zeta)\,
	\left(\wavefunction[k,\mathcal{I}_{cb}](z,\tau,\zeta)\right)^*\;.
\end{equation}
This equation is symmetric under $i \leftrightarrow j$, which implies that
\begin{align}
	Y_{ijk} = Y_{jik}\;.
\end{align}
As we discuss around \cref{eq:exchangesymmetries} in the main text, the
$i\leftrightarrow j$ flip can entail an outer automorphism of the low--energy
gauge symmetry.

\section{Modular transformations of Yukawa couplings}
\label{app:T_transform_yukawa}

In this appendix we will show the different ways in which the Yukawa couplings
obtained from the overlap integrals~\eqref{eq:Yijk} transform under
modular transformations and how that they indeed are modular forms according to
\cref{eq:MetaplecticTrafof}.
 
\subsection{Transformation of the overlap integrals}\label{app:Overlapintegral}

Let us start by discussing how our dictionary
\eqref{eq:WaveFunctionTrafoSummaryT} between the wave functions with torus
parameter $\tau$ and an equivalent torus with parameter $\tau + 1$ allows us  to
infer how the three index Yukawa couplings $Y_{ijk}$ transform. We start with
the $T$ transformation where we use
\eqref{eq:WaveFunctionTrafoSummaryT} for the 2D wave functions.
As we have discussed around \eqref{eq:bparm}, our dictionary involves a shift
of the $z$--coordinate, $\Delta z$. For definiteness we use $\Delta z =
\frac{1}{2}$. Thus, from \cref{eq:Yijk} we have 
\begin{align}\label{eq:YijkTrafoTDetailed}
	Y_{ijk}(\tau+1) &= \int\limits_{\mathds{T}^{2}}\!\D^2 z\,
	\left( \rho(T)_{i, i^{\prime}}^{\psi}
	\mathrm{e}^{\I \pi \mathcal{I}_{ab}\frac{\im z}{\im \tau}}
	\wavefunction[i^{\prime},\mathcal{I}_{ab}]\left( z -\tfrac{1}{2}, \tau , 0 \right)\right)	 \left( \rho(T)_{j, j^{\prime}}^{\psi}\mathrm{e}^{\I \pi \mathcal{I}_{ca}\frac{\im z}{\im \tau}}
	\wavefunction[j^{\prime},\mathcal{I}_{ca}]\left(z -\tfrac{1}{2} , \tau , 0 \right)\right)\notag\\
	 &\quad\cdot\left( \rho(T)_{k, k^{\prime}}^{\psi}\mathrm{e}^{\I \pi \mathcal{I}_{cb}\frac{\im z}{\im \tau}} 
	 \wavefunction[k^{\prime},\mathcal{I}_{cb}] \left(z -\tfrac{1}{2} , \tau , 0 \right)\right)^{*} \notag \\
	&= \int\limits_{\mathds{T}^{2}}\!\D^2 z\,  \mathrm{e}^{\I\pi ({\mathcal{I}}_{ab}+{\mathcal{I}}_{ca}+{\mathcal{I}}_{bc}) \frac{\im z}{2\im \tau}}\, \rho(T)_{i,i^{\prime}}^{\psi}\rho(T)_{j,j^{\prime}}^{\psi}\left(\rho(T)_{k,k^{\prime}}^{\psi}\right)^{*}\notag\\
	&\quad\cdot \wavefunction[i^{\prime},\mathcal{I}_{ab}](z-\tfrac{1}{2},\tau,0)\,
	\wavefunction[j^{\prime},\mathcal{I}_{ca}](z-\tfrac{1}{2},\tau,0)\,
	\left(\wavefunction[k^{\prime},\mathcal{I}_{cb}](z-\tfrac{1}{2},\tau,0)\right)^{*}\;.
\end{align}
Using \cref{eq:FluxCondition}, we find that,
\begin{align}\label{eq:YijkTrafoTDetailed2}
	\MoveEqLeft Y_{ijk}(\tau+1)= \rho(T)_{i,i^{\prime}}^{\psi}\rho(T)_{j,j^{\prime}}^{\psi}\left(\rho(T)_{k,k^{\prime}}^{\psi}\right)^{*}\, \notag\\
	&\quad\cdot \int\limits_{\mathds{T}^{2}}\!\D^2 z\,
	\wavefunction[i^{\prime},\mathcal{I}_{ab}](z-\tfrac{1}{2},\tau,0)\,
	\wavefunction[j^{\prime},\mathcal{I}_{ca}](z-\tfrac{1}{2},\tau,0)\,
	\left(\wavefunction[k^{\prime},\mathcal{I}_{cb}](z-\tfrac{1}{2},\tau,0)\right)^* \;.
\end{align}
We can now define $w:= z-\tfrac{1}{2}$. Then $\D^2 z = \D^2 w$, i.e.\ the
integration measure for torus coordinates and the domain of integration remains
invariant. Thus we find that
\begin{align}\label{eq:YijkTrafoTDetailed3}
	Y_{ijk}(\tau+1)&=\rho(T)_{i,i^{\prime}}^{\psi}\rho(T)_{j,j^{\prime}}^{\psi}\left(\rho(T)_{k,k^{\prime}}^{\psi}\right)^{*}  \int_{\mathbb{T}^{2}}\D^2 w \, \wavefunction[i^{\prime},\mathcal{I}_{ab}](w,\tau,0)\,
	\wavefunction[j^{\prime},\mathcal{I}_{ca}](w,\tau,0)\,
	\left(\wavefunction[k^{\prime},\mathcal{I}_{cb}](w,\tau,0)\right)^* \notag \\
	&= \rho(T)_{i,i^{\prime}}^{\psi}\rho(T)_{j,j^{\prime}}^{\psi}\left(\rho(T)_{k,k^{\prime}}^{\psi}\right)^{*}\, Y_{ijk}(\tau) \notag \\
	&= \mathrm{e}^{\I \pi (i^2/\mathcal{I}_{ab} + j^2/\mathcal{I}_{ca}-k^2/\mathcal{I}_{cb} + i + j - k)}\, Y_{ijk}(\tau)
	\;.
\end{align}
Thus the $z$--dependent phase appearing in our dictionary for $T$ 
transformation~\eqref{eq:oddMshift} cancels out due to the condition
\eqref{eq:FluxCondition}.

For the $S$ transformation of $Y_{ijk}$, we use
\cref{eq:WaveFunctionTrafoSummaryS}, which gives
\begin{align}\label{eq:YijkTrafoSDetailed}
	Y_{ijk}(-1/\tau) &=  \int\limits_{\mathds{T}^{2}}\!\D^2 z\left(-\left( \frac{-\tau }{|\tau |}\right)^{\nicefrac12} \rho(S)_{i, i^{\prime}}^{\psi}\wavefunction[i^{\prime},\mathcal{I}_{ab}] \left( z , \tau , 0 \right)\right) 	 \left( -\left(\frac{-\tau}{|\tau |}\right)^{\nicefrac12}\rho(S)_{j, j^{\prime}}^{\psi}\wavefunction[j^{\prime},\mathcal{I}_{ca}] \left(z , \tau , 0 \right)\right)\notag\\
	 &\quad\cdot\left(-\left(\frac{-\tau}{|\tau |}\right)^{\nicefrac12} \rho(S)_{k, k^{\prime}}^{\psi} \wavefunction[k^{\prime},\mathcal{I}_{cb}] \left(z , \tau , 0 \right)\right)^{*} \notag \\
	 &=-\left(\frac{-\tau}{|\tau |}\right)^{\nicefrac12}\rho(S)_{i, i^{\prime}}^{\psi }\rho(S)_{j, j^{\prime}}^{\psi }\left[\rho(S)_{k, k^{\prime }}^{\psi}\right]^{*}Y_{i^{\prime}j^{\prime}k^{\prime}}\notag\\
	 &=-\left(\frac{-\tau}{|\tau |}\right)^{\nicefrac12} \frac{-\mathrm{e}^{\I\frac{\pi}{4}}}{\sqrt{|\mathcal{I}_{ab}\mathcal{I}_{bc}\mathcal{I}_{bc}|}}\sum_{i^{\prime }=0}^{\mathcal{I}_{ab}-1}\sum_{j^{\prime }=0}^{\mathcal{I}_{ca}-1}\sum_{k^{\prime }=0}^{{\mathcal{I}_{cb}}-1}\mathrm{e}^{2\pi \I \left(\frac{i i^{\prime }}{\mathcal{I}_{ab}}+\frac{jj^{\prime }}{\mathcal{I}_{ca}}+\frac{kk^{\prime }}{\mathcal{I}_{bc}}\right) }Y_{i^{\prime}j^{\prime}k^{\prime}}\;, 
\end{align}
where have used the fact that the automorphy factor and the $\rho(S)^{\psi }$
matrix do not depend in the $z$ coordinate, and then, can be taken out of the
integral. 

\Cref{eq:YijkTrafoTDetailed3,eq:YijkTrafoSDetailed} give the modular
transformations of $Y_{ijk}$. They can be used to infer the possible
modular transformations of the 4D fields.

\subsection[Modular transformation of the lambda-plet of Yukawa couplings]{Modular transformation of the $\rep{\lambda}$--plet of Yukawa couplings}

The $\lambda$--plet of holomorphic Yukawa couplings
\eqref{eq:HolomorphicYukawaCoupling}, $\mathcal{Y}_{ijk}(\tau)=
\vartheta\orb{\widehat{\alpha}_{ijk}/\lambda}{0}(0,\lambda\,\tau)$, transforms
as a modular form of weight $\nicefrac{1}{2}$.  To see this, let us first
investigate how $\mathcal{Y}_{\widehat{\alpha}}(\tau)$, where
$\widehat{\alpha}:=\widehat{\alpha}_{ijk}\in\mathds{Z}_{\lambda}$,  behaves
under $T$. Obviously,
\begin{equation}
 \mathcal{Y}_{\widehat{\alpha}}(\tau) \xmapsto{~T~}
 \mathcal{Y}_{\widehat{\alpha}}(\tau)=\sum\limits_{\ell=-\infty}^\infty
 \exp\left[\I\pi\left(\frac{\widehat{\alpha}}{\lambda}+\ell\right)^2
 \lambda\,(\tau+1)\right]\;.
\end{equation}
The phase can be manipulated to give
\begin{equation}
 \I\pi\left(\frac{\widehat{\alpha}}{\lambda}+\ell\right)^2\lambda\,(\tau+1)
 =
 \I\pi\left(\frac{\widehat{\alpha}}{\lambda}+\ell\right)^2\lambda\,\tau
 +\I\pi\frac{(\widehat{\alpha}+\lambda\,\ell)^2}{\lambda}\;.
\end{equation}
The second term can be rewritten as
\begin{equation}
 \I\pi\frac{(\widehat{\alpha}+\lambda\,\ell)^2}{\lambda}
 =\I\pi\frac{\widehat{\alpha}^2}{\lambda}
 +2\pi\I\,\ell+\I\pi\,\lambda\,\ell^2\;.
\end{equation}
Only the first term on the right--hand side yields a nontrivial phase. The two
others are integer multiples of $2\pi\I$ because $\lambda$ is even. Therefore,
\begin{equation}\label{eq:T_transformation_Yukawa_theta}
 \mathcal{Y}_{\widehat{\alpha}}(\tau+1)=
 \mathrm{e}^{\I\pi\frac{\widehat{\alpha}^2}{\lambda}}\,
 \mathcal{Y}_{\widehat{\alpha}}(\tau)\;.
\end{equation}
Likewise, under $S$
\begin{align}
 \mathcal{Y}_{\widehat{\alpha}}(\tau) \xmapsto{~S~}
 \mathcal{Y}_{\widehat{\alpha}}(-1/\tau)=
 \vartheta\orb{\widehat{\alpha}/\lambda}{0}\bigl(0,-\lambda/\tau\bigr)=:
 \vartheta\orb{\widehat{\alpha}/\lambda}{0}\bigl(0,-1/t\bigr)
\end{align}
where $t:=\tau/\lambda$. Then
\begin{align}
 \MoveEqLeft\mathcal{Y}_{\widehat{\alpha}}(-1/\tau)= 
 \sqrt{-\I\,t}\,\vartheta\orb{0}{\widehat{\alpha}/\lambda}\bigl(0,t\bigr) 
 = 
 \frac{\sqrt{-\I\,\tau}}{\sqrt{\lambda}}\,
 \sum\limits_{\widehat{\beta}=0}^{\lambda-1}
 \mathrm{e}^{\frac{2\pi\I\,\widehat{\alpha}\,\widehat{\beta}}{\lambda}}\,
 \vartheta\orb{\widehat{\beta}/\lambda}{0}\bigl(0,\lambda\,\tau\bigr)
 \notag\\
 &=\left(-\tau\right)^{\nicefrac12}\,
 \sum\limits_{\widehat{\beta}=0}^{\lambda-1}
 \frac{\mathrm{e}^{\frac{\pi \I}{4}}}{\sqrt{\lambda}}
 \mathrm{e}^{\frac{2\pi\I\,\widehat{\alpha}\,\widehat{\beta}}{\lambda}}\
 \vartheta\orb{\widehat{\beta}/\lambda}{0}\bigl(0,\lambda\,\tau\bigr)
 =-\left(-\tau \right)^{\nicefrac12} 
 \,\sum\limits_{\widehat{\beta}=0}^{\lambda-1}
 \left(-\frac{\mathrm{e}^{\frac{\pi \I}{4}}}{\sqrt{\lambda}}\right)\,
 \mathrm{e}^{\frac{2\pi\I\,\widehat{\alpha}\,\widehat{\beta}}{\lambda}}
 \mathcal{Y}_{\widehat{\beta}}(\tau)\;.
 \label{eq:S_transformation_Yukawa_theta}
\end{align}
Here we used~\cref{eq:Poisson,eq:ThetaModularS}. This shows that the
$\lambda$--plet of $\mathcal{Y}_{\widehat{\alpha}}(\tau)$ picks up the correct
automorphy factors to be a modular form of weight $\nicefrac{1}{2}$. Note that
we choose the minus sign in \cref{eq:S_transformation_Yukawa_theta},
anticipating that these transformations comply  with
\cref{eq:MetaplecticTrafof}, for $\varphi(S,\tau)=-\sqrt{-\tau}$, and thus 
with~\cref{eq:tildeGamma_4N-presentation}. Therefore, from
\cref{eq:T_transformation_Yukawa_theta,eq:S_transformation_Yukawa_theta} we get
the representations of the $\lambda$--plet of Yukawa couplings
\eqref{eq:Yukawas_Trafos_L_T}, which we recast here
\begin{subequations}\label{eq:Yukawas_Trafos_L_appendix}
\begin{align}
 \rho_{\rep{\lambda}}(\widetilde S)_{\widehat{\alpha}\widehat{\beta}}&=-\frac{\mathrm{e}^{\I\pi/4}}{\sqrt{\lambda}}
 \,\exp\left(\frac{2\pi\I\,\widehat{\alpha}\,\widehat{\beta}}{\lambda}\right)\;,
 \label{eq:Yukawas_Trafos_L_S_appendix}\\
 \rho_{\rep{\lambda}}(\widetilde T)_{\widehat{\alpha}\widehat{\beta}}&=\exp\left(\frac{\I\pi\,\widehat{\alpha}^2}{\lambda}\right)\,
 \delta_{\widehat{\alpha}\widehat{\beta}}\;.\label{eq:Yukawas_Trafos_L_T_appendix}
\end{align}
\end{subequations}
Finally, although these matrices may be not be irreducible for some choice of
$\mathcal{I}_{\alpha\beta}$, in \cref{sec:Models} we get the irreducible
representation matrix in each case (cf.\ e.g.\ \cref{eq:rho_Y_6,eq:rho_Y_6to4}).
Therefore, \cref{eq:MetaplecticTrafof} is satisfied and the Yukawa couplings
given by \cref{eq:Yijk} are modular forms of weight $k_{Y}=\nicefrac12$. Furthermore, as
discussed in \cref{sec:Models}, the representation matrix will correspond to a
representation of the metaplectic group $\widetilde{\Gamma}_{2\lambda}$, which
implies that the Yukawa couplings have level $2\lambda$.

\begin{acronym}
  \acro{QFT}{quantum field theory}
  \acro{SB}{symmetry based}
  \acro{SM}{Standard Model}
  \acro{SUSY}{supersymmetry}
  \acro{TB}{torus based}
  \acro{UV}{ultraviolet}
\end{acronym}

\providecommand{\bysame}{\leavevmode\hbox to3em{\hrulefill}\thinspace}
\frenchspacing
\newcommand{\origttfamily}{}
\let\origttfamily=\ttfamily
\renewcommand{\ttfamily}{\origttfamily \hyphenchar\font=`\-}

\end{document}